\documentclass[final,12pt,nopreprintline,longtitle]{elsarticle}

\usepackage{amssymb}
\usepackage{amsmath}
\usepackage{booktabs}
\usepackage{subcaption}
\usepackage{csquotes}
\usepackage{xcolor}  

\begin{document}

\begin{frontmatter}



\title{Bayesian Aneurysm Growth Detection via Surface Displacement Modeling}

\author[label1]{Jorge A. Roa Castro\fnref{eq}}
\author[label2]{Abhishek Singh\fnref{eq}}
\author[label2]{Atharva Hans\fnref{eq}}
\author[label1]{Kostiantyn Kondratiuk}
\author[label3]{David Saloner}
\author[label2]{Vitaliy L. Rayz}
\author[label2]{Pavlos P. Vlachos}
\author[label2]{Ilias Bilionis\corref{cor1}}

\cortext[cor1]{Corresponding author.}

\ead{ibilion@purdue.edu}

\affiliation[label1]{organization={Weldon School of Biomedical Engineering, Purdue University},
            addressline={206 S Martin Jischke Dr},
            city={West Lafayette},
            postcode={47907},
            state={Indiana},
            country={USA}}

\affiliation[label2]{organization={School of Mechanical Engineering, Purdue University},
            addressline={585 Purdue Mall},
            city={West Lafayette},
            postcode={47907},
            state={Indiana},
            country={USA}}

\affiliation[label3]{organization={Department of Radiology and Biomedical Imaging, University of California},
            addressline={505 Parnassus Ave},
            city={San Francisco},
            postcode={94143},
            state={California},
            country={USA}}

\fntext[eq]{These authors contributed equally to this work.}

\begin{abstract}

Clinical decisions for unruptured intracranial aneurysms often depend on detecting growth on follow-up magnetic resonance angiography (MRA). Growth is typically judged from manual 2D diameters on a few slices, which vary across clinicians and frequently miss subtle 3D change. Even with 3D segmentations, apparent differences can reflect resolution, segmentation, surface processing, or registration mismatch rather than true growth; most criteria remain heuristic and binary.
We show that a Bayesian displacement-based model using the surrounding vessel as an internal reference achieves strong discrimination of aneurysm growth (AUC  0.86--0.87) and improves agreement with expert labels (Cohen’s \(\kappa\) up to 0.66 vs. 0.35 for volumetric criteria), while providing calibrated posterior probabilities with uncertainty bounds. The method registers baseline and follow-up surfaces, computes normal-directed displacements, and summarizes change as the difference between mean aneurysm displacement and mean displacement on the surrounding non-aneurysmal vessel segment. The vessel segment serves as an internal control for imaging and processing variability, assuming negligible structural change over the surveillance interval. We evaluate two cohorts spanning time-of-flight and contrast-enhanced longitudinal MRA studies: a public dataset labeled from neuroradiologist-provided measurements and an institutional dataset labeled by senior (neurologist) and junior (general physician) raters. Performance is preserved when training on lower-expertise labels, indicating robustness to label variability. Calibrated probabilities may aid clinical decision-making by identifying borderline cases, where high uncertainty can motivate repeat imaging when scan quality or processing variability may explain apparent change. This framework provides interpretable probabilistic growth assessment from longitudinal MRA, reduces dependence on clinician expertise, and supports cross-centre surveillance across scanners and angiography sequences.

\end{abstract}


\end{frontmatter}



\section{Introduction}
\label{sec1}
Aneurysm growth is one of the most powerful predictors of rupture \cite{inoue_annual_2012,mehan_unruptured_2014,van_der_kamp_risk_2021}. Growing intracranial aneurysms (IA) are 30 times more likely to rupture than stable ones \cite{brinjikji_risk_2016}. While most large IAs are intervened, small ($\leq$ 7 mm) are often selected for longitudinal monitoring due to their low risk of rupture in relation with the risk of procedural complications \cite{etminan_unruptured_2016,algra_procedural_2019}. In this setting, growth evidence is commonly the trigger for preventive treatment \cite{thompson_guidelines_2015,etminan_european_2022}. With the increasing availability of imaging modalities, earlier detections, and more patients selected for longitudinal monitoring, an increasing number of management decisions are based primarily on the evidence of growth \cite{etminan_unruptured_2016}.  

Because of its association with rupture, growth has also been proposed as a surrogate endpoint in risk prediction models \cite{backes_phases_2015-1,backes_elapss_2017}. However, there is no consensus on how growth should be determined. In routine clinical practice, growth is typically assessed using manual two-dimensional measurements with electronic calipers on a small number of image slices. These measurements are subject to substantial intra- and inter-observer variability \cite{kim_intraobserver_2017,timmins_reliability_2021,planinc_assessing_2024}, and threshold-based definitions built on continuous measurements remain heterogeneous across studies \cite{malhotra_growth_2017}.

Research workflows increasingly assess growth from co-registered three-dimensional (3D) surface models derived from segmentations of computed tomographic angiography (CTA) or magnetic resonance angiography (MRA) \cite{boussel_aneurysm_2008,firouzian_intracranial_2012,liu_volumetric_2021,bizjak_aneurysm_2024}. These representations enable visualization and quantification of longitudinal changes in aneurysm size and morphology, and they also support downstream analyzes such as computational fluid dynamics (CFD) simulations in patient-specific geometries\cite{goudarzian_predicting_2025}. However, the measured change between two reconstructed surfaces reflects not only biological growth but also cumulative variability introduced by the full processing pipeline. Image resolution and partial-volume effects limit geometric fidelity, and imperfect timing of contrast injection can further degrade image quality. Acquisition-specific artifacts (notably saturation effects in Time-of-Flight (TOF) MRA, and more generally in motion-affected scans) can blur vessel boundaries \cite{maki_effects_1996,tsuruda_artifacts_1992}. Segmentation depends on method choice and parameter settings, including intensity thresholds that can systematically expand or contract lumen geometry or \enquote{melt} angles in bifurcating vessels and aneurysm necks. Surface extraction (often based on Marching Cubes \cite{10.1145/37402.37422}) can leave \enquote{staircasing} artifacts without sufficient smoothing, while smoothing can further alter local curvature and volume. Finally, rigid registration ignores natural vessel position drift, whereas non-rigid registration can inadvertently distort morphology.

Several methods attempt to quantify growth from longitudinal alignment. Firouzian et al.\ \cite{firouzian_intracranial_2012} introduced a groupwise non-rigid registration approach for CTA time series and demonstrated improved agreement with clinical reports compared with independently segmenting each scan, while enabling visualization of local wall displacements. Bizjak and \v{S}piclin proposed a more sophisticated non-rigid registration and morphing strategy for CTA and MRA surface models and derived deformation-based biomarkers with good specificity for growth detection \cite{bizjak_aneurysm_2024}. However, these approaches do not explicitly model uncertainty arising from segmentation, surface modeling, and registration.

A complementary strategy is the volumetric change criterion proposed by Liu et al.\ \cite{liu_volumetric_2021}, which first ensures stability of a reference vessel (volume change $<2\%$) in order to mitigates global bias from threshold selection. Growth is then declared when aneurysm volume increases by $\geq 11\%$. Because the criterion compares volumes rather than surface displacements, it is less sensitive to registration error. In practice, however, it can be labor intensive, requiring repeated segmentation and meshing until the reference-vessel constraint is satisfied.

In this work, we use a non-aneurysmal vessel segment as an internal reference within a displacement-based growth assessment. Each longitudinal surface is partitioned into an aneurysmal segment (aneurysm and adjacent parent vessel) and a healthy-vessel segment (remaining vasculature), which is assumed to undergo negligible change over the surveillance interval. We summarize growth by the difference between mean normal-directed displacements in the two segments and map this patient-level statistic to a posterior probability of growth, with uncertainty bounds, using a Bayesian soft-threshold model.

\section{Methods}
\label{sec:methods}

\subsection{Overview}
\begin{figure}[htbp]
  \centering
  \includegraphics[width=\linewidth]{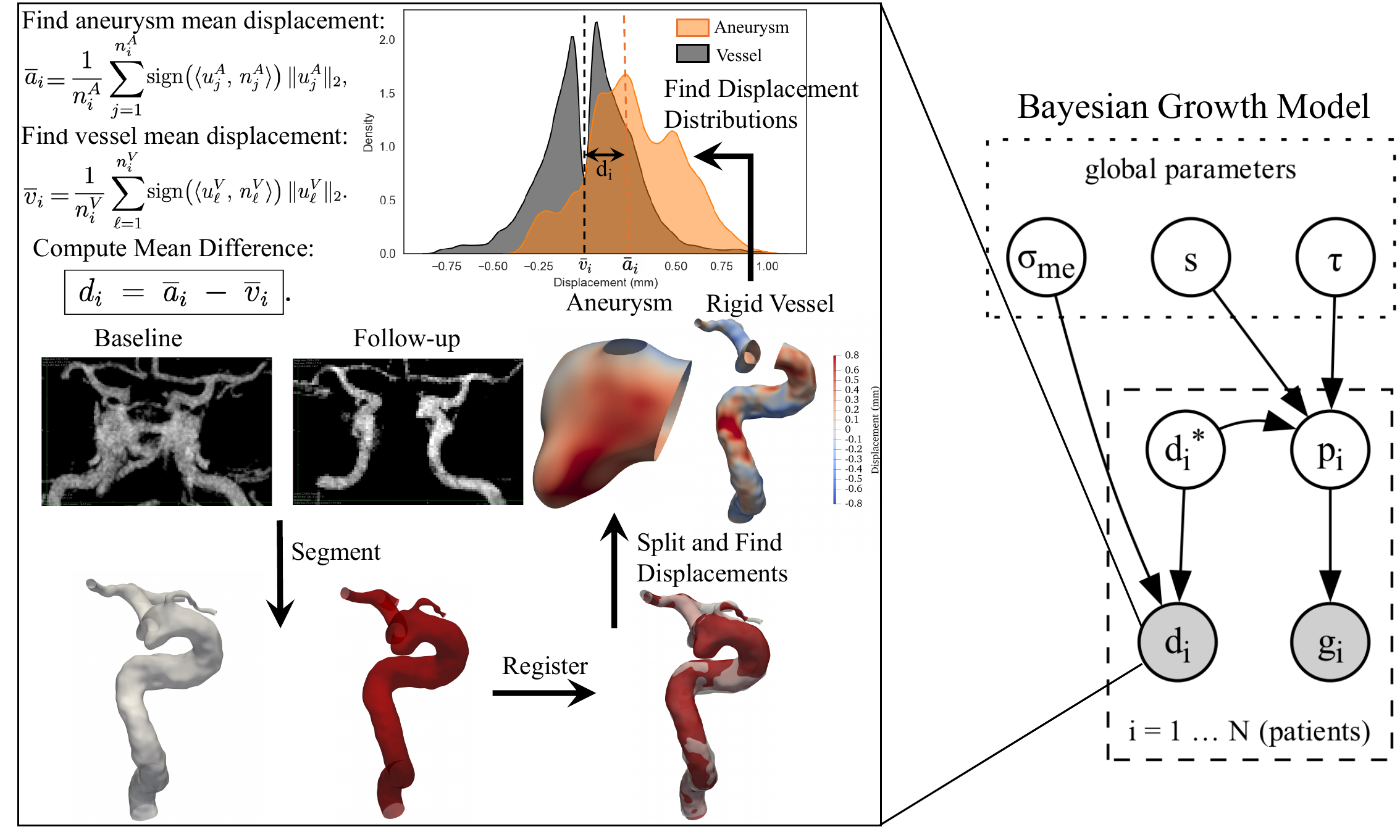}
    \caption{Workflow. Baseline and follow-up surfaces are segmented, the follow-up mesh is rigidly registered to the baseline mesh, and each surface is partitioned into aneurysmal and healthy-vessel segments. Segment-matched nearest-neighbour normal-directed displacements are computed on the baseline vertices (mm). The patient-level mean-shift is defined as $d_i=\overline{a}_i-\overline{v}_i$, the difference between mean displacements on the aneurysmal and healthy-vessel segments. The standardized statistic is modeled with measurement-error standard deviation $\sigma_{\mathrm{me}}$, and a logistic soft-threshold with cut-off $\tau$ and slope $s$ yields $p_i=\Pr(g_i=1)$, the posterior probability of growth.}

  \label{fig:flowchart}
\end{figure}

Our method converts baseline and follow-up surface models from longitudinal magnetic resonance angiography into a posterior probability of aneurysm growth (Figure~\ref{fig:flowchart}).

We segment the vasculature at both time points, generate watertight triangular meshes, and rigidly register the follow-up mesh to the baseline mesh so that changes are evaluated in a common frame (mm). On the baseline mesh, we define an aneurysmal segment (sac and adjacent parent vessel) and a healthy-vessel segment (remaining vasculature).

We establish vertex-wise correspondence between the baseline surface and the registered follow-up surface within each segment and compute per-vertex displacements. We use normal-directed displacements to quantify local change because growth direction can vary over the aneurysm surface and vector averaging can cancel expansion; we retain outward versus inward change by signing the magnitude using the baseline surface normal.

For subject \(i\), we summarize interval change with a displacement-contrast statistic,
\[
d_i \;=\; \overline{a}_i \;-\; \overline{v}_i,
\]
where \(\overline{a}_i\) and \(\overline{v}_i\) are the mean normal-directed displacements (mm) on the aneurysmal and non-aneurysmal vessel segments, respectively. We map \(d_i\) to a posterior \emph{probability of growth} using a Bayesian soft-threshold model with measurement-error scale \(\sigma_{\mathrm{me}}\), which estimates a cut-off \(\tau\) and slope \(s\) and returns probabilities with credible intervals. Distances are standardized within each training cohort and mapped back to millimetres for reporting; full details follow in the subsequent subsections.

\subsection{Data}
\subsubsection{Imaging}

We analyzed MRA images from two longitudinal cerebral aneurysm cohorts. The institutional cohort was drawn from an IRB-approved surveillance study at the University of California, San Francisco (UCSF), conducted from April 2001 to July 2019. Contrast-enhanced (CE) MRA was acquired using institutional protocols on either a Philips Achieva 1.5~T scanner (in-plane voxel size 0.47~mm, slice thickness 0.7~mm) or a Siemens Skyra 3~T scanner (0.7~mm isotropic). Based on image quality and the availability of longitudinal scans, we included 39 patients with 42 unruptured aneurysms. The median baseline-to-follow-up interval was 1.7 years (range 0.5--9.9 years).

For external validation, we used the open-access Metro North Hospital and Health Service (MNHHS) Time-of-Flight (TOF) MRA dataset \cite{de_nys_time--flight_2024}. We screened 24 patients with follow-up imaging and included 16 patients with 19 aneurysms, based on image quality and the absence of intra-luminal clots or evidence of treatment. When multiple imaging time points were available, we selected a baseline–follow-up pair that maximized image quality and representation of growth events, minimizing class imbalance. The median interval was about a year (range 0.3--8.5 years).

\subsubsection{Clinical growth classification}

We emulated routine clinical practice for growth assessment. For the UCSF cohort, an junior (J.R.C., general physician) and senior (K.K., neurologist) MDs independently measured maximum aneurysm diameters  along the image x-, y-, and z-axes, blinded to all surface-model outputs. We defined growth as a \(\geq 1\) mm increase in any dimension between baseline and follow up images, following recent standardization efforts \cite{hackenberg_definition_2019}. For the MNHHS cohort, we used the neuroradiologist-provided diameter measurements and applied the same 1 mm rule. Because the MNHHS dataset  originally used a 2 mm threshold, our binary labels for MNHHS may differ from those reported in the source release.

\subsection{Surface model preparation}

Both MDs jointly segmented the aneurysm and surrounding vasculature at both time points. For the UCSF cohort, we followed the protocol of Liu et al.\ \cite{liu_volumetric_2021}: we defined a region of interest (ROI) around the aneurysm that included a reference segment of healthy vessel, applied intensity thresholding to separate lumen from background, and manually refined the mask to remove leak artifacts and unattached structures, such as small vessels. We generated watertight triangular meshes using the marching cubes algorithm \cite{10.1145/37402.37422}. We then enforced reference-segment stability by measuring its volume at baseline ($t{=}0$) and follow-up ($t{=}1$); if the volumes differed by more than 2\%, we re-segmented the follow-up scan with an adjusted threshold until the difference was below 2\%. For the MNHHS cohort, we did not apply iterative thresholding; instead, to mirror common practice, thresholds were selected to maximize vessel coverage while avoiding leaks and artifacts.

For each vasculature, we obtained a baseline and follow-up mesh. We rigidly registered the follow-up mesh to the baseline using iterative closest point (ICP) with a point-to-plane objective and applied the resulting transform so that both time points were expressed in the baseline coordinate frame.

Finally, we partitioned each surface into an aneurysmal segment and a healthy-vessel segment using two user-defined cutting planes placed proximal and distal to the aneurysm along the parent vessel. After registration, we reused the same planes for both time points to ensure consistent segmentation boundaries, and all subsequent computations were performed separately on the two segments.

\subsection{Displacement maps on baseline coordinates}

We quantify displacement on the \emph{baseline} surface and express all geometry in the baseline coordinate frame. Let
\[
X^{A}_{0} = \big(x^{A}_{1},\ldots,x^{A}_{n^{A}}\big)^{\!\top}\in\mathbb{R}^{n^{A}\times 3},
\qquad
X^{V}_{0} = \big(x^{V}_{1},\ldots,x^{V}_{n^{V}}\big)^{\!\top}\in\mathbb{R}^{n^{V}\times 3}
\]
denote the ordered baseline vertex coordinates for the aneurysmal and healthy-vessel segments. After rigid registration of the follow-up surface to the baseline, let
\[
Y^{A}_{1} = \big(y^{A}_{1},\ldots,y^{A}_{n^{A}_{1}}\big)^{\!\top}\in\mathbb{R}^{n^{A}_{1}\times 3},
\qquad
Y^{V}_{1} = \big(y^{V}_{1},\ldots,y^{V}_{n^{V}_{1}}\big)^{\!\top}\in\mathbb{R}^{n^{V}_{1}\times 3}
\]
be the registered follow-up vertex coordinates in the baseline frame.

We establish correspondence across time by nearest-neighbour search \emph{within the same anatomical segment}. For each baseline vertex, we identify the nearest follow-up vertex (using a KD-tree for efficient search) and define the index maps
\[
\pi^{A}_{j} \;=\; \arg\min_{k\in\{1,\ldots,n^{A}_{1}\}} \big\|y^{A}_{k}-x^{A}_{j}\big\|_{2},
\qquad
\pi^{V}_{\ell} \;=\; \arg\min_{m\in\{1,\ldots,n^{V}_{1}\}} \big\|y^{V}_{m}-x^{V}_{\ell}\big\|_{2}.
\]
We then compute displacement vectors (mm),
\[
u^{A}_{j} \;=\; y^{A}_{\pi^{A}_{j}} - x^{A}_{j},
\qquad
u^{V}_{\ell} \;=\; y^{V}_{\pi^{V}_{\ell}} - x^{V}_{\ell}.
\]
To encode inward versus outward change, we use outward unit normals on the baseline surface, \(n^{A}_{j}\) and \(n^{V}_{\ell}\), and define normal-directed displacements
\[
r^{A}_{j} \;=\; \operatorname{sign}\!\big(\langle u^{A}_{j},\, n^{A}_{j}\rangle\big)\,\|u^{A}_{j}\|_{2},
\qquad
r^{V}_{\ell} \;=\; \operatorname{sign}\!\big(\langle u^{V}_{\ell},\, n^{V}_{\ell}\rangle\big)\,\|u^{V}_{\ell}\|_{2}.
\]
Thus, outward movement relative to baseline contributes positively and inward movement negatively, while the magnitude quantifies the local amount of change. We do not use the normal projection $\langle u, n\rangle$ as the displacement magnitude. While $\langle u, n\rangle$ is a natural choice when small deformations are known to occur primarily along the surface normal, we do not assume that the apparent baseline-to-follow-up correspondence displacement is strictly normal to the baseline surface. Using the projection as the sole magnitude could therefore attenuate cases where outward remodeling is present but not aligned with the baseline normal everywhere. Instead, we use the baseline normal only to assign direction (inward versus outward), and retain the Euclidean displacement length to provide a conservative, geometry-agnostic measure of local change.

We summarize each segment by the mean normal-directed displacement for case \(i\in\{1,\ldots,N\}\),
\[
\overline{a}_{i} \;=\; \frac{1}{n^{A}_{i}}\sum_{j=1}^{n^{A}_{i}} r^{A}_{ij},
\qquad
\overline{v}_{i} \;=\; \frac{1}{n^{V}_{i}}\sum_{\ell=1}^{n^{V}_{i}} r^{V}_{i\ell},
\]
and define a case-level mean-shift,
\[
d_{i} \;=\; \overline{a}_{i} - \overline{v}_{i}\, .
\]
Here \(\overline{v}_{i}\) serves as an internal reference for acquisition- and processing-related bias. Values near zero indicate similar apparent movement in the two segments, whereas \(d_{i}>0\) indicates greater outward change concentrated in the aneurysmal segment. We use \(d_i\) as the input to the Bayesian classifier.

\subsection{From distances to probabilities: Bayesian soft-threshold model}
\label{sec:bst}

The input to the classifier is the scalar mean-shift \(d_i\in\mathbb{R}\) (mm). Our aim is to infer the probability that the aneurysm grew between baseline and follow-up while propagating uncertainty in \(d_i\). Let
\[
g_i \in \{0,1\}
\]
denote the observed clinical growth label, where \(g_i=1\) indicates growth by the clinical criterion and \(g_i=0\) indicates no growth. We model \(g_i\) as a Bernoulli outcome whose success probability increases monotonically with an error-corrected distance. The corresponding probabilistic graphical model is shown in Figure~\ref{fig:flowchart}.

\textbf{Standardization.}
To improve numerical stability and to express the threshold and slope on a common, unitless scale, distances are standardized within each training cohort:
\[
\tilde d_i \;=\; \frac{d_i-\mu}{\sigma},\qquad
\mu=\frac{1}{N}\sum_{i=1}^N d_i,\quad
\sigma=\sqrt{\frac{1}{N}\sum_{i=1}^N (d_i-\mu)^2}.
\]
All model parameters are defined on the standardized scale. Quantities are mapped back to millimetres using \(d=\sigma\,\tilde d+\mu\).

\textbf{Generative model with measurement error.}
The standardized distance \(\tilde d_i\) is treated as a noisy observation of a latent, error-corrected standardized distance \(d_i^{\ast}\):
\[
d_i^{\ast} \sim \mathcal{N}(0,1),\qquad
\tilde d_i \mid d_i^{\ast},\sigma_{\mathrm{me}} \sim \mathcal{N}\!\bigl(d_i^{\ast},\sigma_{\mathrm{me}}^2\bigr),
\]
where \(\sigma_{\mathrm{me}}>0\) represents variability in the measured distance induced by acquisition and processing. This layer ensures that uncertainty in the distance propagates to uncertainty in the inferred growth probability.

\textbf{Soft-threshold likelihood for clinical growth labels.}
Conditioned on \(d_i^{\ast}\), the probability of clinical growth is defined by a logistic soft threshold,
\[
p_i \;=\; \Pr(g_i=1 \mid d_i^{\ast},\tau,s)
\;=\; \operatorname{logistic}\!\bigl(s\,(d_i^{\ast}-\tau)\bigr),
\qquad
\operatorname{logistic}(x)=\frac{1}{1+e^{-x}},
\]
and the observed label is modeled as
\[
g_i \mid p_i \sim \mathrm{Bernoulli}(p_i).
\]
The link is bounded in \([0,1]\), increases monotonically with \(d_i^{\ast}\), and approaches a hard threshold as \(s\to\infty\). The cut-off \(\tau\) is the standardized distance where \(p_i=0.5\). The slope \(s>0\) controls how rapidly the probability transitions near \(\tau\).

\textbf{Priors.}
Weakly informative priors are placed on the standardized scale:
\[
\tau \sim \mathcal{N}(0,1),\qquad
s \sim \mathrm{HalfNormal}(5),\qquad
\sigma_{\mathrm{me}} \sim \mathrm{HalfCauchy}(0.5).
\]
These choices reflect the interpretation and scale of each parameter. Since standardization centers distances at zero, \(\tau\sim\mathcal N(0,1)\) places the 50\% point near the cohort mean while allowing several standard deviations of variation. The half-normal prior on \(s\) enforces monotonicity and favors moderate transition steepness on the standardized scale, while avoiding heavy tails that can destabilize sampling. The half-Cauchy prior on \(\sigma_{\mathrm{me}}\) enforces positivity while remaining flexible for a scale parameter, allowing the data to support larger measurement variability when warranted.

\textbf{Posterior inference.}
Let \(\theta=\{\tau,s,\sigma_{\mathrm{me}}\}\) and \(d^{\ast}=\{d_i^{\ast}\}_{i=1}^N\). The joint posterior is
\[
p(\theta,d^{\ast} \mid \tilde d, g) \;\propto\;
\Big[\prod_{i=1}^N p(g_i \mid d_i^{\ast},\tau,s)\,p(\tilde d_i \mid d_i^{\ast},\sigma_{\mathrm{me}})\,p(d_i^{\ast})\Big]\;p(\theta).
\]
Posterior samples are obtained using Markov chain Monte Carlo with the No-U-Turn Sampler (NUTS) \cite{hoffman_gelman_nuts_2014}. Convergence is assessed using \(\hat R\), effective sample sizes, and trace plots. For larger cohorts or higher-dimensional extensions (e.g., vertex-wise latent fields), variational inference may be used as a scalable approximation while still yielding uncertainty estimates \cite{hans2023stochastic, hans2024bayesian}.

\textbf{Posterior predictive probability for a new case.}
Given a new observed distance \(d_{\text{new}}\) (mm), we compute \(\tilde d_{\text{new}}=(d_{\text{new}}-\mu)/\sigma\) using the training \((\mu,\sigma)\). For each posterior draw \(\theta^{(t)}=\{\tau^{(t)},s^{(t)},\sigma_{\mathrm{me}}^{(t)}\}\), measurement uncertainty is propagated by sampling a latent standardized distance
\(d_{\text{new}}^{\ast}\sim p(d_{\text{new}}^{\ast}\mid \tilde d_{\text{new}},\sigma_{\mathrm{me}}^{(t)})\)
implied by the Gaussian prior and measurement model, and then computing
\[
p_{\text{new}}^{(t)}=\operatorname{logistic}\!\Bigl(s^{(t)}\bigl(d_{\text{new}}^{\ast}-\tau^{(t)}\bigr)\Bigr).
\]
The collection \(\{p_{\text{new}}^{(t)}\}\) provides a Monte Carlo approximation to the posterior predictive distribution of the growth probability, which we summarize by its median and a 95\% highest-density interval (HDI).

\subsection{Training, validation, and evaluation}
We fit one Bayesian soft-threshold model for each reference label set (junior, senior, and external). For a given reference, the training data are pairs \(\{(d_i,g_i)\}_{i=1}^{N}\), where \(d_i\) is the patient-level mean-shift (mm) and \(g_i\in\{0,1\}\) denotes the corresponding binary growth assessment. Model performance is summarized by discrimination (ROC AUC) and agreement with the reference labels after dichotomizing posterior probabilities at 0.5, reported as percentage agreement and Cohen’s \(\kappa\).

\textbf{Within-cohort evaluation.}
Within each cohort, performance is assessed using leave-one-out cross-validation (LOOCV). For each held-out case \(i\), the model is trained on the remaining \(N-1\) cases: distances are standardized using the training fold statistics \((\mu_{-i},\sigma_{-i})\), posterior inference is performed with NUTS, and a posterior predictive growth probability is computed for the held-out case by propagating measurement uncertainty through posterior draws. Repeating this procedure for all cases yields one out-of-sample probability per aneurysm; discrimination and agreement metrics are then computed once from the pooled set of LOOCV out-of-sample predictions (rather than averaged across folds).

\textbf{Cross-cohort and cross-reference evaluation.}
To assess transfer across cohorts and imaging protocols, a model trained on one cohort is applied to the other without refitting. Specifically, distances in the evaluation cohort are standardized using the training cohort statistics \((\mu,\sigma)\); posterior predictive probabilities are then computed under the training posterior by propagating measurement uncertainty using \(\sigma_{\mathrm{me}}\) and the posterior draws of \((\tau,s)\).

Full sampling diagnostics (trace plots, \(\hat R\), effective sample sizes) and posterior predictive checks are reported in the Supplementary Methods.

\subsection{Reference method and computational profile}
As a reference method, we implement the volumetric growth criterion proposed by Liu et al.~\cite{liu_volumetric_2021}, which declares growth when aneurysm volume increases by \(\geq 11\%\) while the reference-vessel volume changes by \(<2\%\). We evaluate its agreement with each set of reference labels and report these results alongside the proposed Bayesian classifier.

Posterior inference with the No-U-Turn Sampler (NUTS) completes in a few seconds on a standard laptop CPU for the cohort sizes considered here, allowing models to be re-fit routinely as additional cases become available.

\section{Results and Discussions}
\label{sec:results}

We first report the resulting labels and inter-rater reliability for diameter-based growth assessment, then interpret cohort-level posterior predictions and within-cohort leave-one-out performance, evaluate cross-cohort transfer and comparison to the published volumetric criterion, present representative cases, and conclude with multi-centre deployment implications, limitations, and future directions.

\subsection{Growth labeling and inter-rater agreement}

In the UCSF cohort (n = 42 aneurysms), the junior rater classified 11 cases (26\%) as growing, whereas the senior rater classified 8 cases (19\%) as growing. In the MNHHS cohort (n = 19 aneurysms), application of the 1 mm growth criterion to the provided measurements identified 6 cases (32\%) as growing. Within the institutional cohort,  agreement for continuous measurements was good to excellent across dimensions at both baseline and follow-up (ICC \(=0.71\)–\(0.94\); Table~\ref{tab:interrater}), using the interpretation of Koo and Li~\cite{koo_guideline_2016}. When these measurements were dichotomized into growth versus no growth using the 1\,mm rule, agreement decreased to moderate (Cohen’s \(\kappa=0.53\)). This reduction is consistent with information loss from thresholding and the sensitivity of a fixed diameter increment to slice selection and partial-volume effects in MRA, and is in line with prior reports of rater variability~\cite{timmins_reliability_2021,planinc_assessing_2024}.

\begin{table}[htbp]
\centering
\caption{Inter-rater agreement for 2D aneurysm diameter measurements and derived growth classification in the institutional dataset. ICC values refer to absolute agreement between raters for individual dimensions at baseline (BL) and follow-up (FU). Cohen’s $\kappa$ quantifies agreement for binary growth classification.}
\label{tab:interrater}
\begin{tabular}{@{}llll@{}}
\toprule
\textbf{Metric} & \textbf{Measurement Context} & \textbf{Value} & \textbf{Interpretation} \\ 
\midrule
Cohen’s $\kappa$ & Growth classification (2D) & 0.53 & Moderate agreement \\
ICC (Width) & Baseline (BL) & 0.80 & Good agreement \\
ICC (Depth) & Baseline (BL) & 0.93 & Excellent agreement \\
ICC (Height) & Baseline (BL) & 0.71 & Good agreement \\
ICC (Width) & Follow-up (FU) & 0.81 & Good agreement \\
ICC (Depth) & Follow-up (FU) & 0.94 & Excellent agreement \\
ICC (Height) & Follow-up (FU) & 0.71 & Good agreement \\
\bottomrule
\end{tabular}
\end{table}

\subsection{Posterior thresholds and within-cohort probabilistic predictions}

\begin{figure}[htbp]
  \centering
  \includegraphics[width=\linewidth]{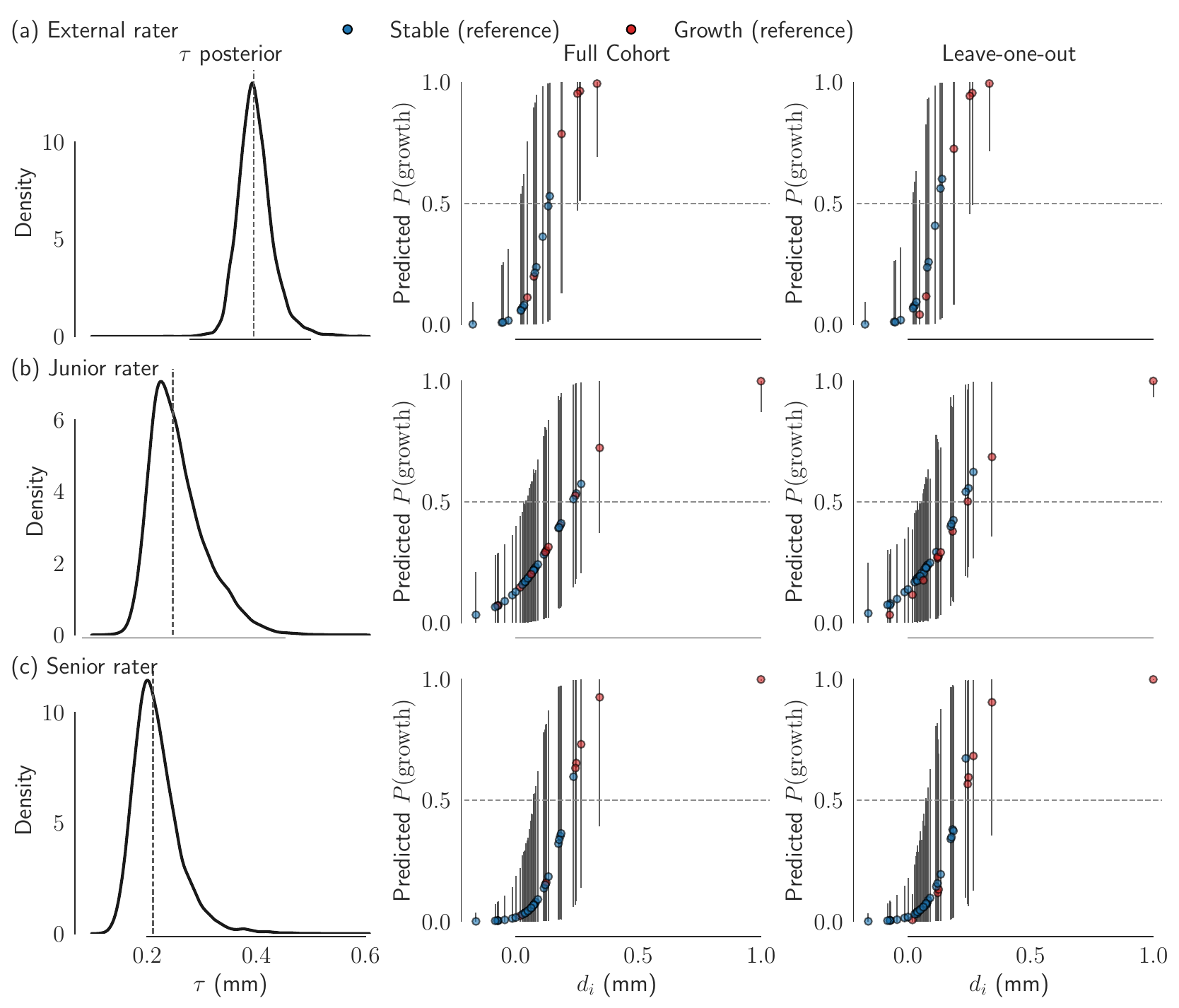}
  \caption{\textbf{Posterior thresholds and within-cohort predictions across raters.}
  Rows correspond to the training reference: (a) external rater (MNHHS), (b) junior rater (UCSF), (c) senior rater (UCSF). \emph{Left}: posterior density of the soft-threshold parameter $\tau$ expressed in millimetres; dashed line marks the posterior median. \emph{Middle}: predicted growth probability $P(\mathrm{growth})$ versus mean-shift $d_i$ (mm) for all cases in the same cohort; points show posterior medians with 95\% highest-density interval (HDI) bars, coloured by the rater’s label (red = growth, blue = stable); dashed horizontal line at $P{=}0.5$ indicates the decision cut-off. \emph{Right}: leave-one-out (LOOCV) predictions displayed analogously.}
  \label{fig:prob_panel}
\end{figure}

Figure~\ref{fig:prob_panel} summarizes the fitted soft-threshold models. Across references, the inferred 50\% point (the cut-off \(\tau\) expressed in millimetres) lies at positive values, consistent with growth presenting as greater outward change on the aneurysm segment relative to the non-aneurysmal vessel segment. Notably, within the UCSF cohort the model trained on junior labels yields a broader posterior for \(\tau\) than the model trained on senior labels, despite similar posterior medians. This difference indicates that label variability is reflected in the learned uncertainty of the transition point, rather than being forced into a single fixed threshold. At the same time, the similarity of posterior medians suggests that the mean-shift \(d_i\) provides a stable summary of differential change across these reference label sets.

Patient-level predictions show the expected monotone relationship between the mean-shift \(d_i\) and posterior growth probability, with the widest credible intervals concentrated around intermediate \(d_i\) values where cases are clinically borderline. At the extremes of \(d_i\), posterior probabilities concentrate near 0 or 1, indicating confident predictions when differential change is clearly absent or clearly present. Leave-one-out predictions follow the same overall pattern (Fig.~\ref{fig:prob_panel}, right), indicating that the learned mapping from \(d_i\) to probability is stable under refitting on nearby training subsets.

\subsection{Discrimination ability and agreement with reference}

\begin{table}[ht]
\centering
\caption{Discrimination (AUC) and agreement (Cohen’s $\kappa$) between the Bayesian classifier and reference growth labels. Values outside parentheses use the full fit; values in parentheses are leave-one-out cross-validation (LOOCV). Evaluation references in bold indicate the rater used for training. The volumetric rule~\cite{liu_volumetric_2021} is shown for comparison.}
\label{tab:kappa_results}
\begin{tabular}{@{}l l l c c@{}}
\toprule
\textbf{Model} & \textbf{Evaluation reference} & \textbf{Dataset} & \textbf{AUC} & \textbf{Cohen's $\kappa$} \\
\midrule
\textbf{Junior Model}
  & \textbf{Junior}& Internal & 0.71 (0.66) & 0.21 (0.21) \\
  & Senior& Internal & 0.86 (0.83) & 0.66 (0.66) \\
  & External& Public   & 0.87 (0.82) & 0.58 (0.51) \\
\textbf{Senior Model}
  & Junior& Internal & 0.71 (0.66) & 0.21 (0.21) \\
  & \textbf{Senior}& Internal & 0.86 (0.83) & 0.66 (0.66) \\
  & External& Public   & 0.87 (0.82) & 0.58 (0.51) \\
\textbf{External Model}
  & Junior& Internal & 0.72 (0.66) & 0.26 (0.21) \\
  & Senior& Internal & 0.86 (0.83) & 0.39 (0.66) \\
  & \textbf{External}& Public & 0.87 (0.82) & 0.51 (0.51) \\
\midrule
\textbf{Volumetric}
  & Junior& Internal & 0.73         & 0.46        \\
  & Senior& Internal & 0.71         & 0.35        \\
  & External& Public   & 0.72         & 0.38        \\
\bottomrule
\end{tabular}
\end{table}

\begin{figure}
  \centering
  \includegraphics[width=\linewidth]{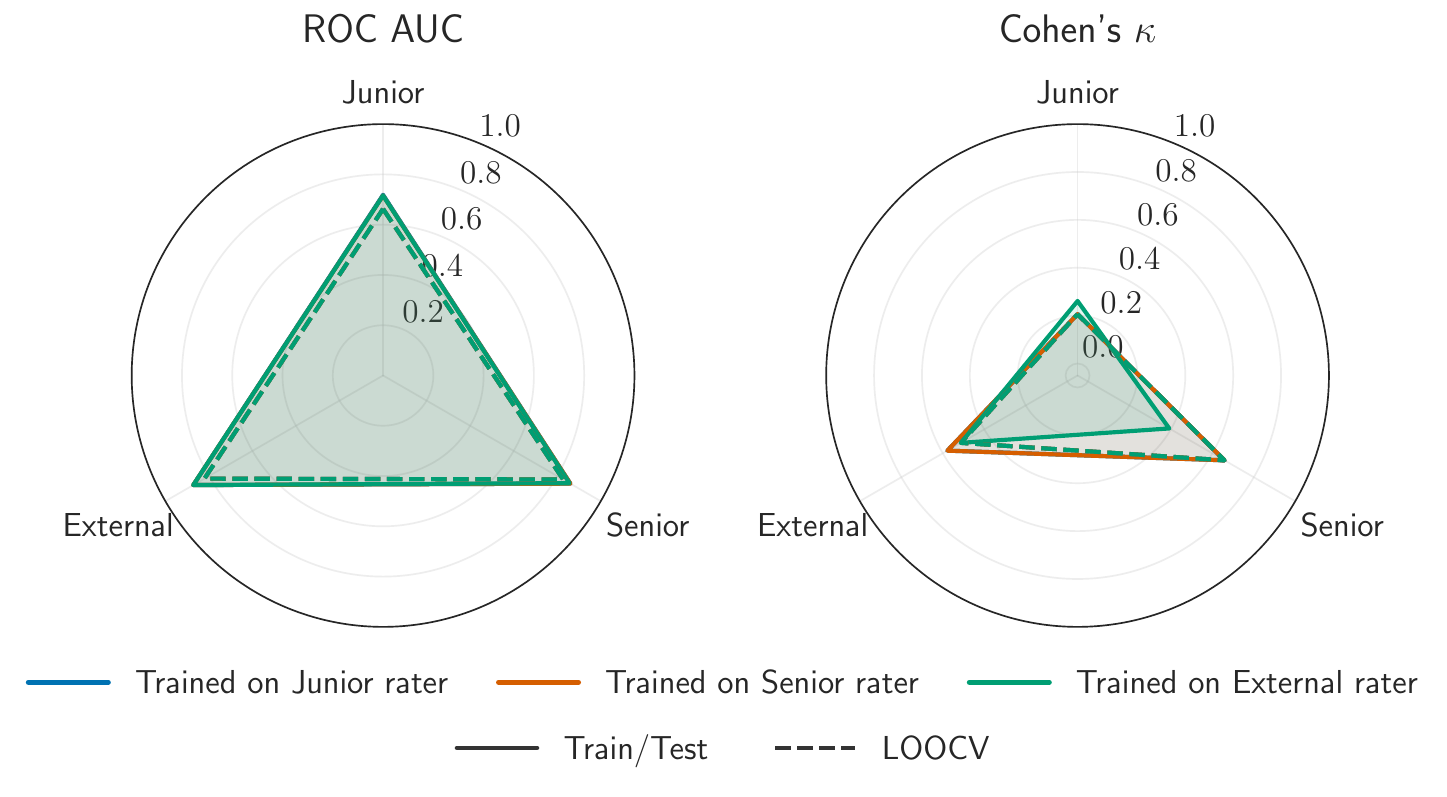}
  \caption{Models performance against labels from all three raters. AUC and Cohen’s $\kappa$, with LOOCV performance in dotted lines.}
  \label{fig:radar}
\end{figure}

Table~\ref{tab:kappa_results} and Fig.~\ref{fig:radar} summarize discrimination and agreement against each reference label set. Discrimination is quantified by the area under the receiver operating characteristic curve (ROC AUC), and agreement is summarized by Cohen’s \(\kappa\) after dichotomizing posterior probabilities at 0.5. 

Within the institutional cohort, the two models trained on different local references (junior versus senior) yield numerically identical ROC AUC and \(\kappa\) values when evaluated against any fixed reference label set. We verified that this is not a production error but reflects identical case-level predictions. This behavior arises because both models are trained on the same underlying continuous predictor (the mean-shift \(d_i\)) and differ only in the binary labels used to estimate the logistic mapping. Given the limited sample size and substantial overlap between rater labels, the fitted decision functions converge to effectively equivalent thresholds over the observed range of \(d_i\). As a result, posterior probabilities induce identical rankings (driving identical AUC) and identical classifications at the 0.5 threshold (driving identical \(\kappa\)). This convergence indicates that model performance is primarily determined by the underlying displacement-derived feature rather than the specific choice of rater labels, and suggests robustness of the learned mapping to moderate label variability.

Relative to the volumetric criterion of Liu et al.~\cite{liu_volumetric_2021}, the Bayesian model aligns substantially better with the senior reference: AUC increases by \(0.15\) from \(0.71\) to \(0.86\) and \(\kappa\) increases by \(0.31\) from \(0.35\) to \(0.66\) (\(\approx\!+88\%\)). In contrast, agreement with the junior reference is higher for the volumetric criterion (AUC \(0.73\) versus \(0.71\); \(\kappa\) \(0.46\) versus \(0.21\)), consistent with greater variability in these labels and with the fact that a rule-based volumetric threshold can mirror a noisier binary reference more closely without necessarily improving discrimination against senior-expert labels.

Cross-cohort evaluation maintains performance at a level comparable to within-cohort results (Table~\ref{tab:kappa_results}). This is non-trivial because the cohorts differ in scanners and acquisition protocols (time-of-flight MRA in MNHHS versus contrast-enhanced MRA in UCSF), segmentation practice, and labeling conventions. Despite these sources of heterogeneity, the learned soft-threshold and measurement-error components yield a consistent probability mapping when distances are standardized using the training cohort statistics. We return to implications for multi-center use in a later subsection.

\subsection{Representative patient cases}

\begin{figure}[htbp]
  \centering
  \includegraphics[width=\linewidth]{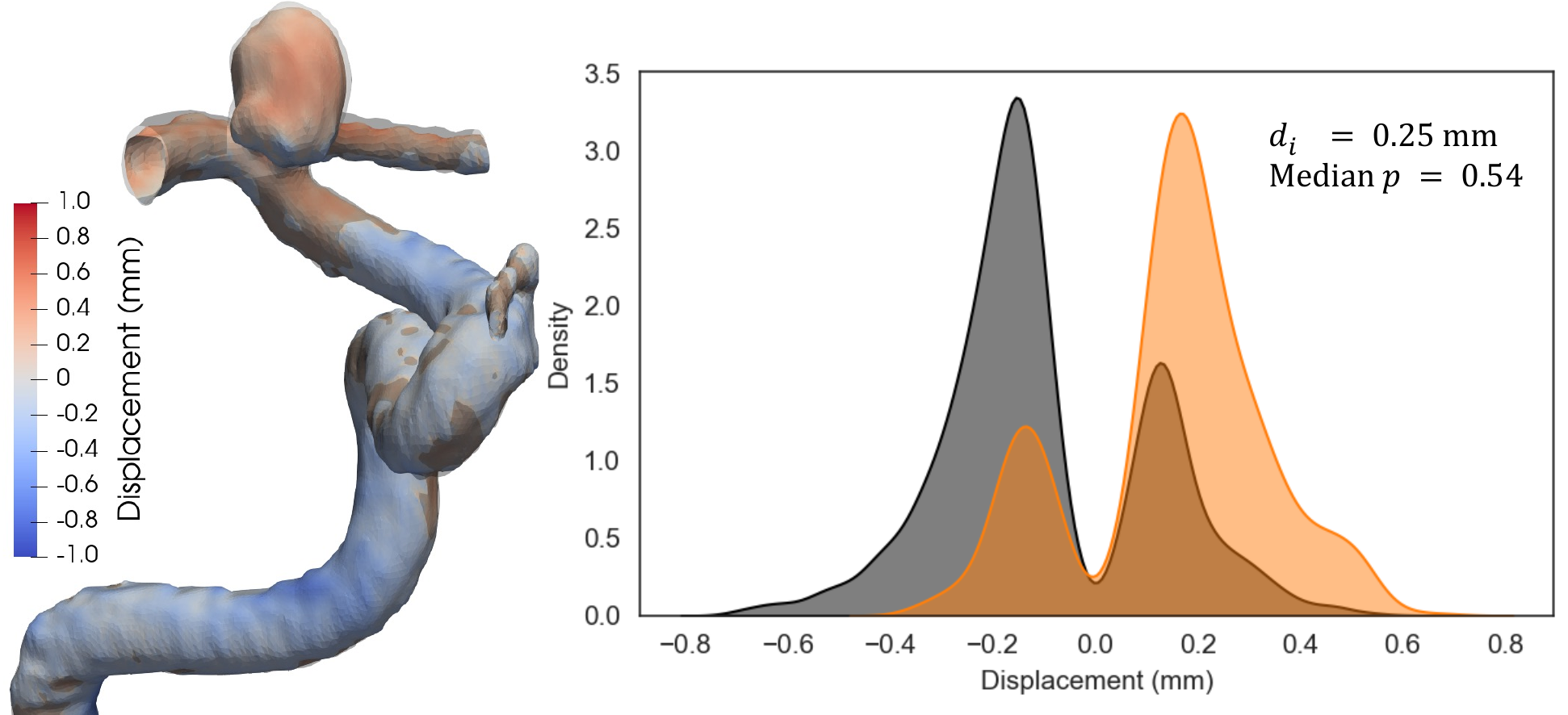}
    \caption{Representative growing internal carotid artery (ICA) bifurcation aneurysm. Left: normal-directed displacement on the baseline surface; registered follow-up surfaces are overlaid (ANY, translucent white; VES, translucent black). Right: kernel density estimates of the normal-directed displacement distributions for ANY (orange) and VES (black), annotated with the Bayesian model’s median growth probability.}
  \label{fig2:mnhhs}
\end{figure}

\begin{figure}[htbp]
  \centering

  \begin{subfigure}[t]{\linewidth}
    \centering
    \includegraphics[width=\linewidth]{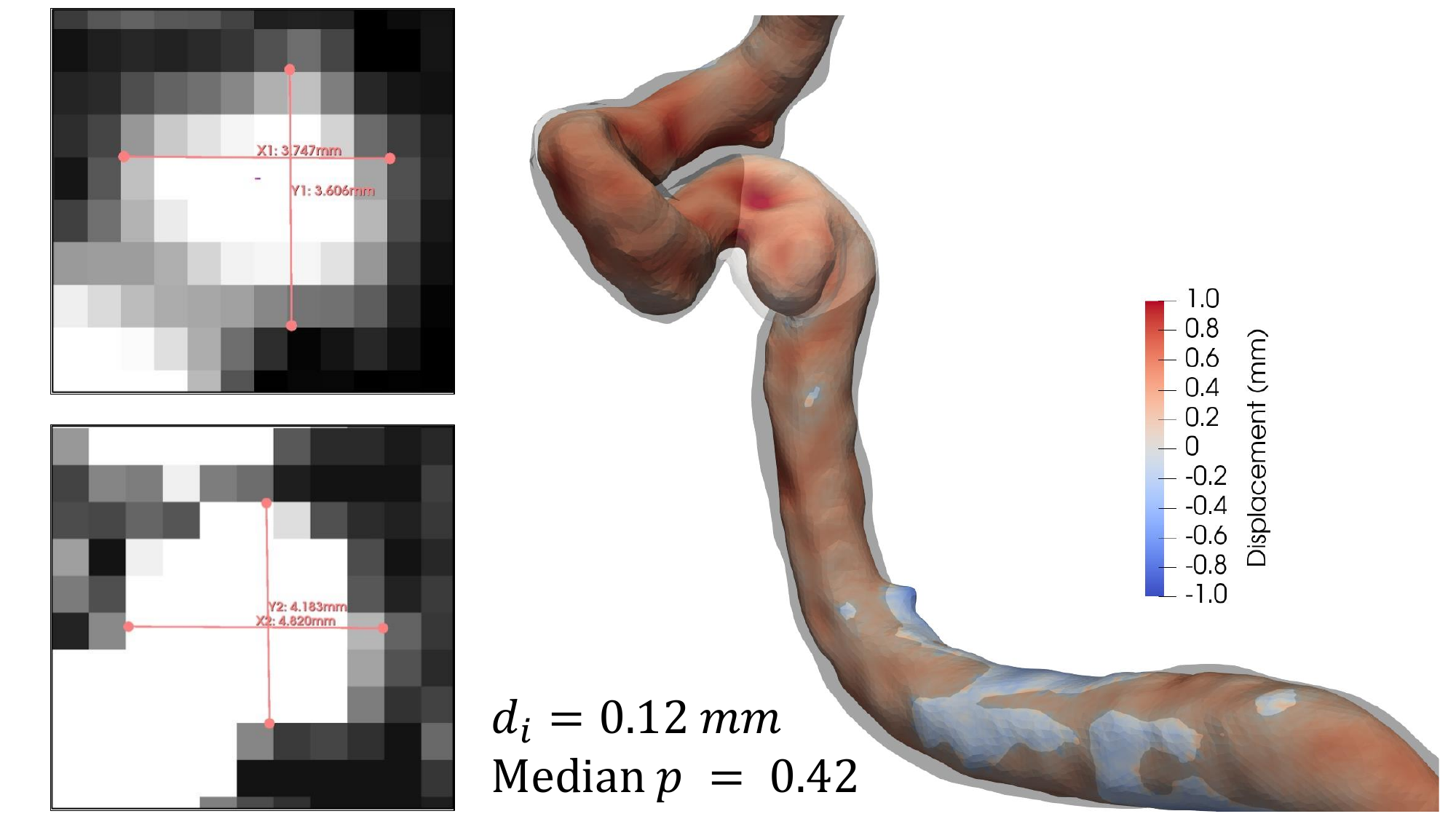}
    \caption{}
    \label{fig_3A:ucsf_case_1}
  \end{subfigure}

  \vspace{0.6em}

  \begin{subfigure}[t]{\linewidth}
    \centering
    \includegraphics[width=\linewidth]{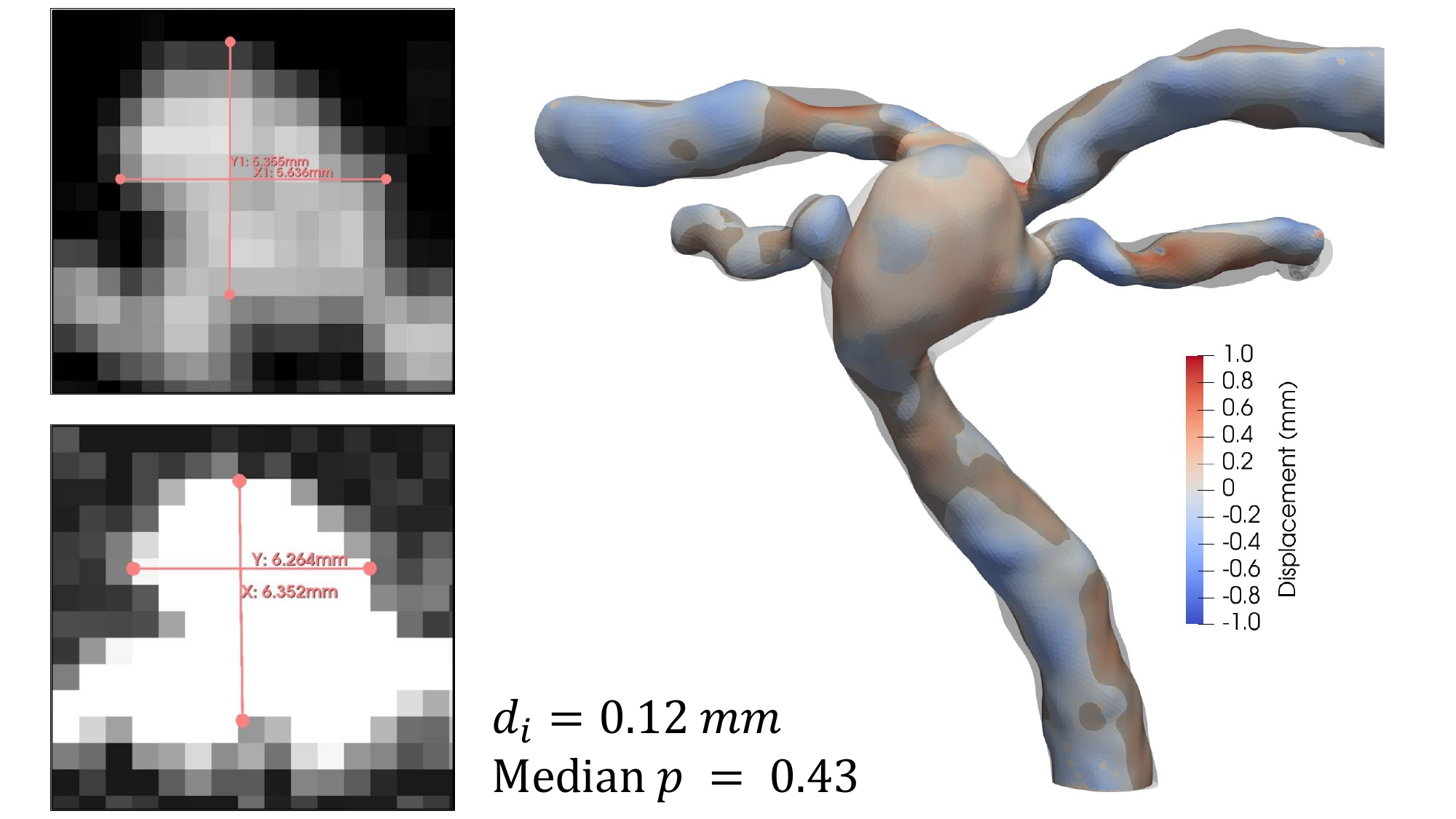}
    \caption{}
    \label{fig_3B:ucsf_case_2}
  \end{subfigure}

  \caption{Representative UCSF cases. Left: (top) baseline 2D cross-section with manual caliper measurements; (bottom) follow-up cross-section with matched calipers. Right: baseline 3D surface coloured by normal-directed displacement (mm) with registered follow-up surfaces overlaid (ANY, translucent white; VES, translucent black), annotated with the Bayesian model’s \(d_i\) and median posterior growth probability.}
  \label{fig:ucsf_cases}
\end{figure}

Figures~\ref{fig2:mnhhs} and \ref{fig:ucsf_cases} illustrate how the proposed mean-shift statistic \(d_i\) and its posterior growth probability behave in individual subjects, and how these decisions compare with the volumetric rule of Liu et al.~\cite{liu_volumetric_2021}. Across cases, the examples emphasize two practical points: (i) subtracting the healthy-vessel displacement baseline helps distinguish focal aneurysm change from global acquisition/processing drift; and (ii) probability outputs are most informative in borderline cases where rule-based thresholds provide little margin.

Figure~\ref{fig2:mnhhs} shows an internal carotid artery bifurcation aneurysm from the MNHHS cohort. The displacement map indicates a largely shared inward displacement on the non-aneurysmal vasculature while the aneurysm segment exhibits relative outward change. Although the volumetric rule is only marginally satisfied (11.8\% aneurysm-volume increase), the mean-shift is clearly positive (\(d_i=0.25\)~mm), leading to a posterior median growth probability of 0.54 and agreement with the senior label. This case illustrates the intended role of the healthy-vessel reference: it absorbs scan- and processing-related drift that appears across the surface and highlights the differential change localized to the aneurysm segment.

Figure~\ref{fig:ucsf_cases} presents two institutional cases that highlight common failure modes of threshold-based assessment. In the ICA lateral aneurysm (Fig.~\ref{fig_3A:ucsf_case_1}), both segments show a broadly positive displacement shift, suggesting a global outward bias rather than isolated aneurysm expansion. Accordingly, the differential statistic remains modest (\(d_i=0.12\)~mm) and the posterior median probability is 0.42, leading to a stable classification that agrees with the senior rater and disagrees with the junior rater. This example illustrates how the internal vessel reference can improve specificity when apparent change is driven by cohort- or scan-specific effects rather than focal aneurysm deformation.

In the basilar artery aneurysm (Fig.~\ref{fig_3B:ucsf_case_2}), segmentation and registration are particularly challenging due to multiple nearby branch vessels and complex local geometry. Although both clinicians labeled growth, the volumetric increase is below the published threshold (8.7\%), and the mean-shift remains small (\(d_i=0.12\)~mm), yielding a posterior median probability of 0.43 (stable). The displacement map suggests that the largest apparent outward changes are concentrated near the branching region rather than presenting as a coherent focal expansion of the aneurysm dome. This pattern is consistent with known difficulties in reconstructing bifurcation geometry under limited resolution, where local angles can be \enquote{melted} and small segmentation inconsistencies can produce localized apparent displacements. In such settings, probabilistic outputs provide a transparent indication of ambiguity, motivating closer review or repeat imaging when clinical concern remains high.

\subsection{Harmonizing cross-centre measurement heterogeneity}
\label{sec:harmonization}

\begin{figure}[htbp]
  \centering
  \includegraphics[width=\linewidth]{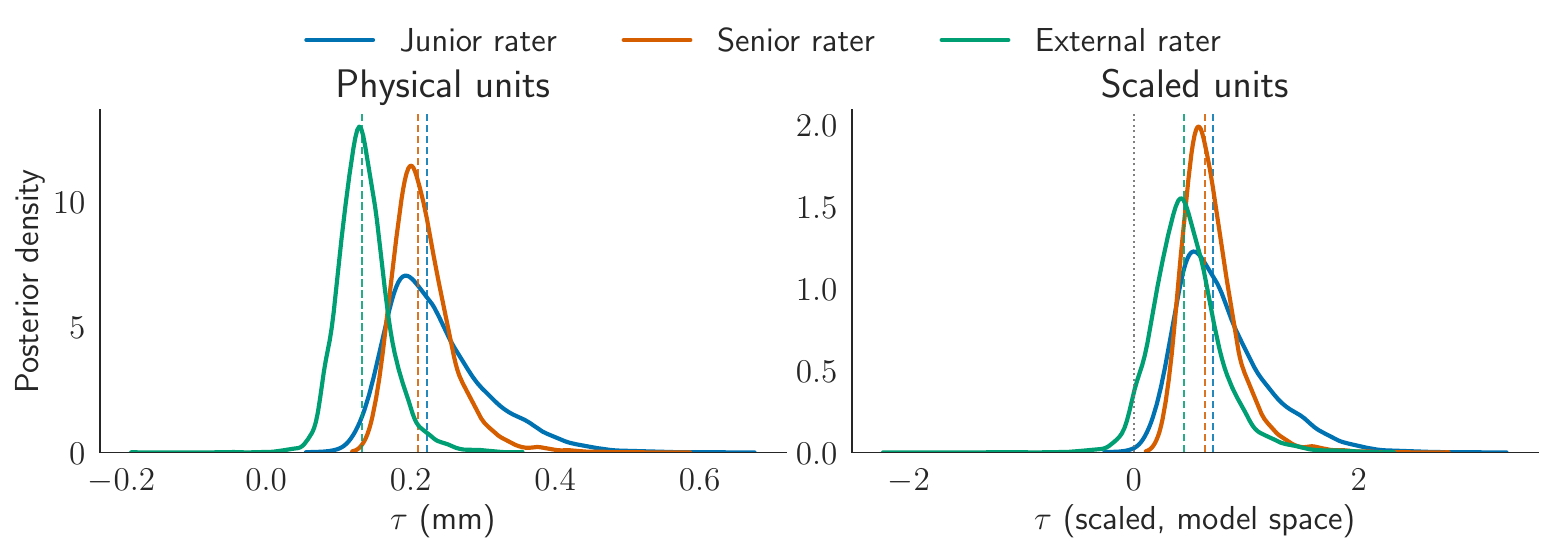}
  \caption{\textbf{Posterior soft-threshold across sites and scales.} Posteriors of the cut-off parameter \(\tau\) for models trained on external rater (MNHHS, time-of-flight MRA), senior rater (UCSF, contrast-enhanced MRA), and physician (UCSF, contrast-enhanced MRA) reference labels. \emph{Left:} \(\tau\) expressed in millimetres. \emph{Right:} the same posteriors on the standardized scale \(z=(d-\mu)/\sigma\). Dashed lines indicate posterior medians.}
  \label{fig:posterior-tau}
\end{figure}

The cross-cohort results can be interpreted by examining the learned cut-off \(\tau\), i.e., the distance at which the logistic link assigns a 50\% growth probability (Fig.~\ref{fig:prob_panel}). Figure~\ref{fig:posterior-tau} reports the posterior of \(\tau\) both in millimetres and on the standardized scale used for inference, \(z=(d-\mu)/\sigma\). In physical units, the posteriors differ across cohorts: the external rater-trained model (MNHHS time-of-flight MRA) exhibits a narrower distribution and a lower median cut-off than the two UCSF models (contrast-enhanced MRA), consistent with systematic differences in spatial resolution and contrast mechanism. 

After cohort-wise standardization by the empirical mean \(\mu\) and standard deviation \(\sigma\) of the mean-shift distances, the \(\tau\) posteriors overlap substantially (Fig.~\ref{fig:posterior-tau}, right). This convergence indicates that the model is primarily learning a decision rule on a relative distance scale, while the cohort-specific location and spread of \(d_i\) absorb scanner- and protocol-dependent shifts. Standardization harmonizes heterogeneous distance distributions without modifying the core likelihood or requiring site-specific tuning.

This form of harmonization is not a guarantee of domain invariance: it assumes that the sources of variability captured by \(\mu\), \(\sigma\), and the inferred measurement-error scale remain comparable to those represented in the training data. Substantial departures in acquisition quality, segmentation practice, or registration behavior may therefore require recalibration (e.g., refitting \(\mu\), \(\sigma\), and \(\sigma_{\mathrm{me}}\) on a small local set) before deployment. We discuss these limitations and practical implications next.

\subsection{Limitations and Future Directions}

\paragraph{Validation and reference labels}
This study is constrained by cohort size and class imbalance, which limit the precision with which the soft threshold and measurement-error scale can be estimated. Reference labels are derived from routine two-dimensional (2D) diameter measurements on magnetic resonance angiography (MRA), which are known to be sensitive to slice selection and partial-volume effects and can vary across raters. In the institutional cohort, rater experience and the inherent coarseness of the 1\,mm decision rule likely introduce label noise that bounds achievable agreement, irrespective of the underlying model. In the public cohort, the provided measurements were not originally generated for a 1\,mm rule, and applying a different thresholding convention than the dataset’s original release may further increase label mismatch. Finally, surface models in the institutional cohort were generated jointly by the two raters under a standardized protocol, but inter-operator reliability of geometry generation was not assessed; in the public cohort, surfaces were generated by a single operator using pragmatic threshold selection, which improves realism but reduces direct comparability across cohorts.

A natural methodological extension is to explicitly model rater- and site-level variability via hierarchical priors, enabling partial pooling across annotators and acquisition settings while preserving rater-specific thresholds and noise scales~\cite{hans2020quantifying,hans2023bayesian}.

\paragraph{Assumptions behind the internal reference}
The method assumes that the non-aneurysmal vessel segment undergoes negligible structural change over the surveillance interval so that its measured displacements primarily reflect acquisition- and processing-related variability. Slow biological changes in vessel caliber have been reported~\cite{bullitt_effects_2010}, and focal pathology (e.g., plaque) could violate this assumption in some patients. The framework also assumes that segmentation and registration uncertainty affect aneurysmal and non-aneurysmal segments similarly. This may be imperfect: flow-related signal loss is more common within aneurysms (particularly in time-of-flight MRA), and limited spatial resolution can differentially degrade sharp bifurcation geometry relative to smoother parent vessels. We mitigated these effects by focusing on small aneurysms typical of surveillance and by applying minimal smoothing sufficient to remove staircasing, but residual differential bias cannot be excluded.

\paragraph{Registration, partitioning, and correspondence}
Rigid registration error is not uniform across the cerebrovasculature; distal branches can exhibit larger misalignment due to motion, smaller caliber, and reduced influence on a global alignment objective. We reduced this effect by centering models on the aneurysm and restricting spatial extent, but residual drift contributes to dispersion in the distance–risk relationship. Anatomical partitioning requires manual cut-plane placement; inevitable inclusion of small amounts of healthy vessel within the aneurysmal segment is conservative, as it tends to reduce the apparent difference between segments.

Correspondence is established by within-segment nearest-neighbour mapping from baseline vertices to the registered follow-up surface. This choice does not track material points and implicitly treats change as occurring along shortest geometric paths; heterogeneous or curved remodeling trajectories cannot be recovered from two time points. Inter-scan intervals also vary across patients, so the same displacement can reflect different underlying growth rates. The measurement-error term absorbs part of this variability, but it does not resolve these structural limitations.

\paragraph{Model form and outputs}
We deliberately use a logistic soft threshold to map a single scalar distance to a growth probability, keeping inference stable for modest sample sizes. With larger and more heterogeneous datasets, this functional form may be too restrictive, motivating more flexible link functions or feature expansions. In addition, the current output is a calibrated \emph{global} probability of growth; it does not provide statistical evidence for localized “hot spots” of remodeling. A principled extension is a spatially resolved model that returns vertex-level posterior growth probabilities by benchmarking aneurysm displacements against local variability on adjacent vessel wall. Such maps could also serve as a scaffold for multimodal analysis by testing whether regions of elevated growth probability co-localize with adverse hemodynamic metrics derived from CFD, 4D Flow MRI, or particle image/tracking velocimetry (PIV/PTV)~\cite{boussel_aneurysm_2008,brindise_multi-modality_2019,goudarzian_predicting_2025, hans2025smurf}.

\paragraph{Practical considerations and deployment}
The method requires high-quality surface models at multiple time points and accurate separation of aneurysm and non-aneurysmal vessel. In the present pipeline, segmentation and aneurysm isolation remain the most labor-intensive steps and are potential sources of user variability. While automated approaches exist, integration into routine workflows remains limited~\cite{hsu_survey_2025}. Future work should therefore evaluate end-to-end performance with automated segmentation and partitioning.

\section{Conclusions}
We presented an interpretable Bayesian framework for detecting intracranial aneurysm growth from longitudinal MRA using co-registered 3D surface models. The method summarizes interval change with a displacement \emph{contrast}: the difference between mean normal-directed displacements on the aneurysm segment and on an adjacent non-aneurysmal vessel segment. By using the vessel segment as an internal reference, the approach partially absorbs systematic effects from segmentation, meshing, and residual registration, and returns a posterior \emph{probability of growth} with credible uncertainty bounds rather than a binary decision.

Across two cohorts acquired with different angiography sequences and labeled by raters of varying expertise, the model achieved strong discrimination against senior-expert references and preserved performance under cross-cohort transfer after cohort-wise standardization. Agreement with clinician-assigned labels is bounded by the intrinsic variability of diameter-based assessment. Crucially, when trained on junior labels, the model maintained similar agreement with the senior reference as senior-trained models, while representing label inconsistency as increased posterior uncertainty rather than a shifted decision boundary. This uncertainty is clinically actionable: borderline scans with elevated uncertainty can be flagged for closer review or repeat imaging when apparent change may plausibly be explained by measurement variability.

This work establishes a foundation for quantitative, uncertainty-aware aneurysm surveillance from longitudinal clinical imaging. Future efforts will (i) extend the global classifier to vertex-wise posterior growth maps to localize remodeling and support focal interpretation; (ii) incorporate hierarchical structure to model rater- and site-dependent thresholds and noise scales within a unified framework; and (iii) integrate automated segmentation, cross-time alignment, and probabilistic inference into a streamlined pipeline to reduce operator dependence. We will also evaluate multimodal extensions by co-registering growth-probability maps with hemodynamic descriptors (e.g., wall shear stress and oscillatory shear index from 4D Flow MRI, and PIV/PTV- or CFD-derived fields) to test mechanistic hypotheses of aneurysm instability.

\section*{Funding}

Support for this research was provided by the National Institutes of Health (NIH), National Heart, Lung, and Blood Institute (NHLBI) under grant R01 HL115267. Ilias Bilionis and Atharva Hans were supported by the National Science Foundation (grant 2347472).

\section*{Declaration of competing interest}
The authors declare no competing interests.

\section*{Declaration of generative AI and AI-assisted technologies in the manuscript preparation process}
During the preparation of this work the authors used ChatGPT in order to assist with LaTeX formatting. After using this tool, the authors reviewed and edited the content as needed and take full responsibility for the content of the published article.




  \bibliographystyle{elsarticle-num} 
  \bibliography{arxiv_refs}

@article{inoue_annual_2012,
  title = {Annual rupture risk of growing unruptured cerebral aneurysms detected by magnetic resonance angiography: {Clinical} article},
  volume = {117},
  doi = {10.3171/2012.4.JNS112225},
  number = {1},
  journal = {Journal of Neurosurgery},
  author = {Inoue, Takashi and Shimizu, Hiroaki and Fujimura, Miki and Saito, Atsushi and Tominaga, Teiji},
  month = jul,
  year = {2012},
  pages = {20--25}
}

@article{mehan_unruptured_2014,
  title = {Unruptured intracranial aneurysms conservatively followed with serial {CT} angiography: could morphology and growth predict rupture?},
  volume = {6},
  doi = {10.1136/neurintsurg-2013-010944},
  number = {10},
  journal = {Journal of NeuroInterventional Surgery},
  author = {Mehan, William A. and Romero, Javier M. and Hirsch, Joshua A. and Sabbag, David J. and Gonzalez, Ramon G. and Heit, Jeremy J. and Schaefer, Pamela W.},
  month = dec,
  year = {2014},
  pages = {761--766}
}

@article{van_der_kamp_risk_2021,
  title = {Risk of {Rupture} {After} {Intracranial} {Aneurysm} {Growth}},
  volume = {78},
  doi = {10.1001/jamaneurol.2021.2915},
  number = {10},
  journal = {JAMA Neurology},
  author = {van der Kamp, Laura T. and Rinkel, Gabriel J. E. and Verbaan, Dagmar and van den Berg, Ren{\'e} and Vandertop, W. Peter and Murayama, Yuichi and Ishibashi, Toshihiro and Lindgren, Antti and Koivisto, Timo and Teo, Mario and St George, Jerome and Agid, Ronit and Radovanovic, Ivan and Moroi, Junta and Igase, Keiji and van den Wijngaard, Ido R. and Rahi, Melissa and Rinne, Jaakko and Kuhmonen, Johanna and Boogaarts, Hieronymus D. and Wong, George K. C. and Abrigo, Jill M. and Morita, Akio and Shiokawa, Yoshiaki and Hackenberg, Katharina A. M. and Etminan, Nima and van der Schaaf, Irene C. and Zuithoff, Nicolaas P. A. and Vergouwen, Mervyn D. I.},
  month = oct,
  year = {2021},
  pages = {1228--1235}
}

@article{brinjikji_risk_2016,
  title = {Risk {Factors} for {Growth} of {Intracranial} {Aneurysms}: {A} {Systematic} {Review} and {Meta}-{Analysis}},
  volume = {37},
  doi = {10.3174/ajnr.A4575},
  number = {4},
  journal = {AJNR: American Journal of Neuroradiology},
  author = {Brinjikji, W. and Zhu, Y.-Q. and Lanzino, G. and Cloft, H.J. and Murad, M.H. and Wang, Z. and Kallmes, D.F.},
  month = apr,
  year = {2016},
  pages = {615--620}
}

@article{etminan_unruptured_2016,
  title = {Unruptured intracranial aneurysms: development, rupture and preventive management},
  volume = {12},
  doi = {10.1038/nrneurol.2016.150},
  number = {12},
  journal = {Nature Reviews Neurology},
  author = {Etminan, Nima and Rinkel, Gabriel J.},
  month = dec,
  year = {2016},
  pages = {699--713}
}

@article{algra_procedural_2019,
  title = {Procedural {Clinical} {Complications}, {Case}-{Fatality} {Risks}, and {Risk} {Factors} in {Endovascular} and {Neurosurgical} {Treatment} of {Unruptured} {Intracranial} {Aneurysms}: {A} {Systematic} {Review} and {Meta}-analysis},
  volume = {76},
  doi = {10.1001/jamaneurol.2018.4165},
  number = {3},
  journal = {JAMA Neurology},
  author = {Algra, Annemijn M. and Lindgren, Antti and Vergouwen, Mervyn D. I. and Greving, Jacoba P. and van der Schaaf, Irene C. and van Doormaal, Tristan P. C. and Rinkel, Gabriel J. E.},
  month = mar,
  year = {2019},
  pages = {282--293}
}

@article{thompson_guidelines_2015,
  title = {Guidelines for the {Management} of {Patients} {With} {Unruptured} {Intracranial} {Aneurysms}},
  volume = {46},
  doi = {10.1161/STR.0000000000000070},
  number = {8},
  journal = {Stroke},
  author = {Thompson, B. Gregory and Brown, Robert D. and Amin-Hanjani, Sepideh and Broderick, Joseph P. and Cockroft, Kevin M. and Connolly, E. Sander and Duckwiler, Gary R. and Harris, Catherine C. and Howard, Virginia J. and Johnston, S. Claiborne (Clay) and Meyers, Philip M. and Molyneux, Andrew and Ogilvy, Christopher S. and Ringer, Andrew J. and Torner, James and {on behalf of the American Heart Association Stroke Council, Council on Cardiovascular and Stroke Nursing, and Council on Epidemiology and Prevention}},
  month = aug,
  year = {2015},
  pages = {2368--2400}
}

@article{etminan_european_2022,
  title = {European {Stroke} {Organisation} ({ESO}) guidelines on management of unruptured intracranial aneurysms},
  volume = {7},
  doi = {10.1177/23969873221099736},
  number = {3},
  journal = {European Stroke Journal},
  author = {Etminan, Nima and Sousa, Diana Aguiar de and Tiseo, Cindy and Bourcier, Romain and Desal, Hubert and Lindgren, Anttii and Koivisto, Timo and Netuka, David and Peschillo, Simone and L{\'e}meret, Sabrina and Lal, Avtar and Vergouwen, Mervyn DI and Rinkel, Gabriel JE},
  month = jun,
  year = {2022},
  pages = {V}
}

@article{backes_phases_2015-1,
  title = {{PHASES} {Score} for {Prediction} of {Intracranial} {Aneurysm} {Growth}},
  volume = {46},
  doi = {10.1161/STROKEAHA.114.008198},
  number = {5},
  journal = {Stroke},
  author = {Backes, Daan and Vergouwen, Mervyn D.I. and Tiel Groenestege, Andreas T. and Bor, A. Stijntje E. and Velthuis, Birgitta K. and Greving, Jacoba P. and Algra, Ale and Wermer, Marieke J.H. and van Walderveen, Marianne A.A. and terBrugge, Karel G. and Agid, Ronit and Rinkel, Gabriel J.E.},
  month = may,
  year = {2015},
  pages = {1221--1226}
}

@article{backes_elapss_2017,
  title = {{ELAPSS} score for prediction of risk of growth of unruptured intracranial aneurysms},
  volume = {88},
  doi = {10.1212/WNL.0000000000003865},
  number = {17},
  journal = {Neurology},
  author = {Backes, Daan and Rinkel, Gabriel J.E. and Greving, Jacoba P. and Velthuis, Birgitta K. and Murayama, Yuichi and Takao, Hiroyuki and Ishibashi, Toshihiro and Igase, Michiya and terBrugge, Karel G. and Agid, Ronit and J{\"a}{\"a}skel{\"a}inen, Juha E. and Lindgren, Antti E. and Koivisto, Timo and Von Und Zu Fraunberg, Mikael and Matsubara, Shunji and Moroi, Junta and Wong, George K.C. and Abrigo, Jill M. and Igase, Keiji and Matsumoto, Katsumi and Wermer, Marieke J.H. and Van Walderveen, Marianne A.A. and Algra, Ale and Vergouwen, Mervyn D.I.},
  month = apr,
  year = {2017},
  pages = {1600--1606}
}

@article{kim_intraobserver_2017,
  title = {Intraobserver and interobserver variability in {CT} angiography and {MR} angiography measurements of the size of cerebral aneurysms},
  volume = {59},
  doi = {10.1007/s00234-017-1826-y},
  number = {5},
  journal = {Neuroradiology},
  author = {Kim, Hye Jeong and Yoon, Dae Young and Kim, Eun Soo and Lee, Hyung Jin and Jeon, Hong Jun and Lee, Jong Young and Cho, Byung-Moon},
  month = may,
  year = {2017},
  pages = {491--497}
}

@article{timmins_reliability_2021,
  title = {Reliability and {Agreement} of {2D} and {3D} {Measurements} on {MRAs} for {Growth} {Assessment} of {Unruptured} {Intracranial} {Aneurysms}},
  volume = {42},
  doi = {10.3174/ajnr.A7186},
  number = {9},
  journal = {AJNR: American Journal of Neuroradiology},
  author = {Timmins, K.M. and Kuijf, H.J. and Vergouwen, M.D.I. and Otten, M.J. and Ruigrok, Y.M. and Velthuis, B.K. and van der Schaaf, I.C.},
  month = sep,
  year = {2021},
  pages = {1598--1603}
}

@article{planinc_assessing_2024,
  title = {Assessing accuracy and consistency in intracranial aneurysm sizing: human expertise vs. artificial intelligence},
  volume = {14},
  doi = {10.1038/s41598-024-65825-4},
  number = {1},
  journal = {Scientific Reports},
  author = {Planinc, Andrej and {\v{S}}pegel, Nina and Podobnik, Zala and {\v{S}}inigoj, Uro{\v{s}} and Skubic, Petra and Choi, June Ho and Park, Wonhyoung and Robi{\v{c}}, Tina and Tabor, Nika and Jarabek, Leon and {\v{S}}piclin, {\v{Z}}iga and Bizjak, {\v{Z}}iga},
  month = jul,
  year = {2024},
  pages = {16080}
}

@article{malhotra_growth_2017,
  title = {Growth and {Rupture} {Risk} of {Small} {Unruptured} {Intracranial} {Aneurysms}: {A} {Systematic} {Review}},
  volume = {167},
  doi = {10.7326/M17-0246},
  number = {1},
  journal = {Annals of Internal Medicine},
  author = {Malhotra, Ajay and Wu, Xiao and Forman, Howard P. and Grossetta Nardini, Holly K. and Matouk, Charles C. and Gandhi, Dheeraj and Moore, Christopher and Sanelli, Pina},
  month = jul,
  year = {2017},
  pages = {26--33}
}

@article{boussel_aneurysm_2008,
  title = {Aneurysm {Growth} {Occurs} at {Region} of {Low} {Wall} {Shear} {Stress}: {Patient}-{Specific} {Correlation} of {Hemodynamics} and {Growth} in a {Longitudinal} {Study}},
  volume = {39},
  doi = {10.1161/STROKEAHA.108.521617},
  number = {11},
  journal = {Stroke},
  author = {Boussel, Loic and Rayz, Vitaliy and McCulloch, Charles and Martin, Alastair and Acevedo-Bolton, Gabriel and Lawton, Michael and Higashida, Randall and Smith, Wade S. and Young, William L. and Saloner, David},
  month = nov,
  year = {2008},
  pages = {2997--3002}
}

@inproceedings{firouzian_intracranial_2012,
  address = {San Diego, California, USA},
  title = {Intracranial aneurysm growth quantification in {CTA}},
  doi = {10.1117/12.910713},
  author = {Firouzian, Azadeh and Manniesing, Rashindra and Metz, Coert T. and Klein, Stefan and Velthuis, Birgitta K. and Rinkel, Gabriel J. E. and Van Der Lugt, Aad and Niessen, Wiro J.},
  editor = {Haynor, David R. and Ourselin, S{\'e}bastien},
  month = feb,
  year = {2012},
  pages = {831448}
}

@article{liu_volumetric_2021,
  title = {A {Volumetric} {Metric} for {Monitoring} {Intracranial} {Aneurysms}: {Repeatability} and {Growth} {Criteria} in a {Longitudinal} {MR} {Imaging} {Study}},
  volume = {42},
  doi = {10.3174/ajnr.A7190},
  number = {9},
  journal = {American Journal of Neuroradiology},
  author = {Liu, X. and Haraldsson, H. and Wang, Y. and Kao, E. and Ballweber, M. and Martin, A.J. and McCulloch, C.E. and Faraji, F. and Saloner, D. and {for the UCSF Intracranial Aneurysm Monitoring Group}},
  month = sep,
  year = {2021},
  pages = {1591--1597}
}

@article{bizjak_aneurysm_2024,
  title = {Aneurysm growth evaluation and detection: a computer-assisted follow-up {MRA} analysis},
  volume = {14},
  doi = {10.1038/s41598-024-70453-z},
  number = {1},
  journal = {Scientific Reports},
  author = {Bizjak, {\v{Z}}iga and {\v{S}}piclin, {\v{Z}}iga},
  month = aug,
  year = {2024},
  pages = {19609}
}

@article{goudarzian_predicting_2025,
  title = {Predicting {Cerebral} {Aneurysm} {Rupture}},
  volume = {35},
  doi = {10.1016/j.nic.2025.05.002},
  number = {3},
  journal = {Neuroimaging Clinics of North America},
  author = {Goudarzian, Farshid and Kondratiuk, Kostiantyn and Rayz, Vitaliy L.},
  month = aug,
  year = {2025},
  pages = {333--347}
}

@article{maki_effects_1996,
  title = {The effects of time varying intravascular signal intensity and k-space acquisition order on three-dimensional {MR} angiography image quality},
  volume = {6},
  doi = {10.1002/jmri.1880060413},
  number = {4},
  journal = {Journal of Magnetic Resonance Imaging},
  author = {Maki, Jeffrey H. and Prince, Martin R. and Londy, Frank J. and Chenevert, Thomas L.},
  month = jul,
  year = {1996},
  pages = {642--651}
}

@article{tsuruda_artifacts_1992,
  title = {Artifacts associated with {MR} neuroangiography.},
  volume = {13},
  number = {5},
  journal = {American Journal of Neuroradiology},
  author = {Tsuruda, J and Saloner, D and Norman, D},
  month = sep,
  year = {1992},
  pages = {1411}
}

@article{10.1145/37402.37422,
  title = {Marching cubes: {A} high resolution {3D} surface construction algorithm},
  volume = {21},
  doi = {10.1145/37402.37422},
  number = {4},
  journal = {SIGGRAPH Comput. Graph.},
  author = {Lorensen, William E. and Cline, Harvey E.},
  month = aug,
  year = {1987},
  pages = {163--169}
}

@article{de_nys_time--flight_2024,
  title = {Time-of-{Flight} {MRA} of {Intracranial} {Aneurysms} with {Interval} {Surveillance}, {Clinical} {Segmentation} and {Annotations}},
  volume = {11},
  doi = {10.1038/s41597-024-03397-8},
  number = {1},
  journal = {Scientific Data},
  author = {De Nys, Chloe M. and Liang, Ee Shern and Prior, Marita and Woodruff, Maria A. and Novak, James I. and Murphy, Ashley R. and Li, Zhiyong and Winter, Craig D. and Allenby, Mark C.},
  month = may,
  year = {2024},
  pages = {555}
}

@article{hackenberg_definition_2019,
  title = {Definition and {Prioritization} of {Data} {Elements} for {Cohort} {Studies} and {Clinical} {Trials} on {Patients} with {Unruptured} {Intracranial} {Aneurysms}: {Proposal} of a {Multidisciplinary} {Research} {Group}},
  volume = {30},
  doi = {10.1007/s12028-019-00729-0},
  number = {S1},
  journal = {Neurocritical Care},
  author = {Hackenberg, Katharina A. M. and Algra, Ale and Al-Shahi Salman, Rustam and Fr{\"o}sen, Juhana and Hasan, David and Juvela, Seppo and Langer, David and Meyers, Philip and Morita, Akio and Rinkel, Gabriel and Etminan, Nima and {the Unruptured Aneurysms and SAH CDE Project Investigators} and Suarez, Jose I. and Macdonald, R. Loch and Amin-Hanjani, Sepideh and Brown, Robert D. and De Oliveira Manoel, Airton Leonardo and Derdeyn, Colin P. and Etminan, Nima and Keller, Emanuela and LeRoux, Peter D. and Mayer, Stephan and Morita, Akio and Rinkel, Gabriel and Rufennacht, Daniel and Stienen, Martin N. and Torner, James and Vergouwen, Mervyn D. I. and Wong, George K. C. and Bijlenga, Philippe and Ko, Nerissa and McDougall, Cameron G. and Mocco, J. and Murayama, Yuuichi and Werner, Marieke J. H. and Damani, Rahul and Broderick, Joseph and Dhar, Raj and Jauch, Edward C. and Kirkpatrick, Peter J. and Martin, Renee H. and Muehlschlegel, Susanne and Mutoh, Tatsushi and Nyquist, Paul and Olson, Daiwai and Mejia-Mantilla, Jorge H. and Van Der Jagt, Mathieu and Bambakidis, Nicholas and Brophy, Gretchen and Bulsara, Ketan and Claassen, Jan and Connolly, E. Sander and Hoffer, S. Alan and Hoh, Brian L. and Holloway, Robert G. and Kelly, Adam and Nakaji, Peter and Rabinstein, Alejandro and Vajkoczy, Peter and Woo, Henry and Zipfel, Gregory J. and Chou, Sherry and Dor{\'e}, Sylvain and Dumont, Aaron S. and Gunel, Murat and Kasuya, Hidetoshi and Roederer, Alexander and Ruigrok, Ynte and Vespa, Paul M. and Sarrafzadeh-Khorrasani, Asita Simone and Hackenberg, Katharina A. M. and Huston, John and Krings, Timo and Lanzino, Giuseppe and Meyers, Philip M. and Wintermark, Max and Daly, Janis and Ogilvy, Christopher and Rhoney, Denise H. and Roos, Y. B. and Siddiqui, Adnan and Algra, Ale and Fr{\"o}sen, Juhanna and Hasan, David and Juvela, Seppo and Langer, David J. and Salman, Rustam Al-Shahi and Hanggi, Daniel and Schweizer, Tom and Visser-Meily, Johanna and Amos, Liz and Ludet, Christophe and Moy, Claudia and Odenkirchen, Joanne and Ala'i, Sherita and Esterlitz, Joy and Joseph, Kristen and Sheikh, Muniza},
  month = jun,
  year = {2019},
  pages = {87--101}
}

@article{hoffman_gelman_nuts_2014,
  title = {The {No-U-Turn} Sampler: Adaptively Setting Path Lengths in Hamiltonian Monte Carlo},
  author = {Hoffman, Matthew D. and Gelman, Andrew},
  journal = {Journal of Machine Learning Research},
  volume = {15},
  number = {1},
  pages = {1593--1623},
  year = {2014}
}

@inproceedings{hans2023stochastic,
  title = {Stochastic volumetric reconstruction},
  author = {Hans, Atharva and Bhattacharya, Sayantan and Bilionis, Ilias and Vlachos, Pavlos P},
  booktitle = {15th Int. Symp. on Particle Image Velocimetry-ISPIV},
  year = {2023}
}

@article{hans2024bayesian,
  title = {Bayesian reconstruction of 3D particle positions in high-seeding density flows},
  author = {Hans, Atharva and Bhattacharya, Sayantan and Hao, Kairui and Vlachos, Pavlos and Bilionis, Ilias},
  journal = {Measurement Science and Technology},
  volume = {35},
  number = {11},
  pages = {116002},
  year = {2024},
  publisher = {IOP Publishing}
}

@article{koo_guideline_2016,
  title = {A {Guideline} of {Selecting} and {Reporting} {Intraclass} {Correlation} {Coefficients} for {Reliability} {Research}},
  volume = {15},
  doi = {10.1016/j.jcm.2016.02.012},
  number = {2},
  journal = {Journal of Chiropractic Medicine},
  author = {Koo, Terry K. and Li, Mae Y.},
  month = jun,
  year = {2016},
  pages = {155--163}
}

@inproceedings{hans2020quantifying,
  title = {Quantifying individuals' theory-based knowledge using probabilistic causal graphs: a bayesian hierarchical approach},
  author = {Hans, Atharva and Chaudhari, Ashish M and Bilionis, Ilias and Panchal, Jitesh H},
  booktitle = {International Design Engineering Technical Conferences and Computers and Information in Engineering Conference},
  volume = {83921},
  pages = {V003T03A014},
  year = {2020},
  organization = {American Society of Mechanical Engineers}
}

@article{hans2023bayesian,
  title = {A bayesian hierarchical model for extracting individuals' theory-based causal knowledge},
  author = {Hans, Atharva and Chaudhari, Ashish M and Bilionis, Ilias and Panchal, Jitesh H},
  journal = {Journal of Computing and Information Science in Engineering},
  volume = {23},
  number = {3},
  pages = {031011},
  year = {2023},
  publisher = {American Society of Mechanical Engineers}
}

@article{bullitt_effects_2010,
  title = {The effects of healthy aging on intracerebral blood vessels visualized by magnetic resonance angiography},
  volume = {31},
  doi = {10.1016/j.neurobiolaging.2008.03.022},
  number = {2},
  journal = {Neurobiology of Aging},
  author = {Bullitt, Elizabeth and Zeng, Donglin and Mortamet, Benedicte and Ghosh, Arpita and Aylward, Stephen R. and Lin, Weili and Marks, Bonita L. and Smith, Keith},
  month = feb,
  year = {2010},
  pages = {290--300}
}

@article{brindise_multi-modality_2019,
  title = {Multi-modality cerebral aneurysm haemodynamic analysis: \textit{in vivo} {4D} flow {MRI}, \textit{in vitro} volumetric particle velocimetry and \textit{in silico} computational fluid dynamics},
  volume = {16},
  doi = {10.1098/rsif.2019.0465},
  number = {158},
  journal = {Journal of The Royal Society Interface},
  author = {Brindise, Melissa C. and Rothenberger, Sean and Dickerhoff, Benjamin and Schnell, Susanne and Markl, Michael and Saloner, David and Rayz, Vitaliy L. and Vlachos, Pavlos P.},
  month = sep,
  year = {2019},
  pages = {20190465}
}

@article{hans2025smurf,
  title = {SMURF: Scalable method for unsupervised reconstruction of flow in 4D flow MRI},
  author = {Hans, Atharva and Singh, Abhishek and Vlachos, Pavlos and Bilionis, Ilias},
  journal = {arXiv preprint arXiv:2505.12494},
  year = {2025}
}

@article{hsu_survey_2025,
  title = {A survey of intracranial aneurysm detection and segmentation},
  volume = {101},
  doi = {10.1016/j.media.2025.103493},
  journal = {Medical Image Analysis},
  author = {Hsu, Wei-Chan and Meuschke, Monique and Frangi, Alejandro F. and Preim, Bernhard and Lawonn, Kai},
  month = apr,
  year = {2025},
  pages = {103493}
}



\end{document}


\begin{frontmatter}

\title{Supplementary material for ``Bayesian Aneurysm Growth Detection via Surface Displacement Modeling''}

\author[label1]{Jorge A. Roa Castro\fnref{eq}}
\author[label2]{Abhishek Singh\fnref{eq}}
\author[label2]{Atharva Hans\fnref{eq}}
\author[label1]{Kostiantyn Kondratiuk}
\author[label3]{David Saloner}
\author[label2]{Vitaliy L. Rayz}
\author[label2]{Pavlos P. Vlachos}
\author[label2]{Ilias Bilionis\corref{cor1}}

\cortext[cor1]{Corresponding author.}
\ead{ibilion@purdue.edu}
\fntext[eq]{These authors contributed equally to this work.}

\affiliation[label1]{organization={Weldon School of Biomedical Engineering, Purdue University},
            addressline={206 S Martin Jischke Dr},
            city={West Lafayette},
            postcode={47907},
            state={Indiana},
            country={USA}}

\affiliation[label2]{organization={School of Mechanical Engineering, Purdue University},
            addressline={585 Purdue Mall},
            city={West Lafayette},
            postcode={47907},
            state={Indiana},
            country={USA}}

\affiliation[label3]{organization={Department of Radiology and Biomedical Imaging, University of California},
            addressline={505 Parnassus Ave},
            city={San Francisco},
            postcode={94143},
            state={California},
            country={USA}}

\end{frontmatter}

\setcounter{tocdepth}{2}
\tableofcontents
\clearpage

\section{Scope of this supplement}
This supplement provides the model-validation material referenced in the main manuscript for the Bayesian soft-threshold classifier. The main text presents the model formulation, cohort definitions, and primary discrimination and agreement results. Here we provide the convergence diagnostics, posterior summaries, class-prevalence tables, and posterior predictive checks that support those findings.

For brevity, we report detailed MCMC diagnostics for the model fitted to the external MNHHS cohort, which underlies the external-cohort analyses in the main manuscript. We also report class prevalence for all reference label sets because prevalence shapes both the difficulty of the classification task and the interpretation of calibration in modest-sized cohorts.

\section{Supplementary methods}

\subsection{Posterior inference and convergence assessment}
We performed posterior inference with the No-U-Turn Sampler (NUTS) using four chains. To assess convergence, we examined the same diagnostics used throughout the main study workflow: visual inspection of chain traces, split-\(\hat R\), and bulk effective sample size (ESS). We also inspected the one-dimensional posterior marginals to confirm that the fitted threshold, slope, and measurement-error parameters were unimodal and showed no clear pathologies such as chain separation or pronounced multimodality.

\begin{figure}[H]
\centering
\includegraphics[width=\linewidth]{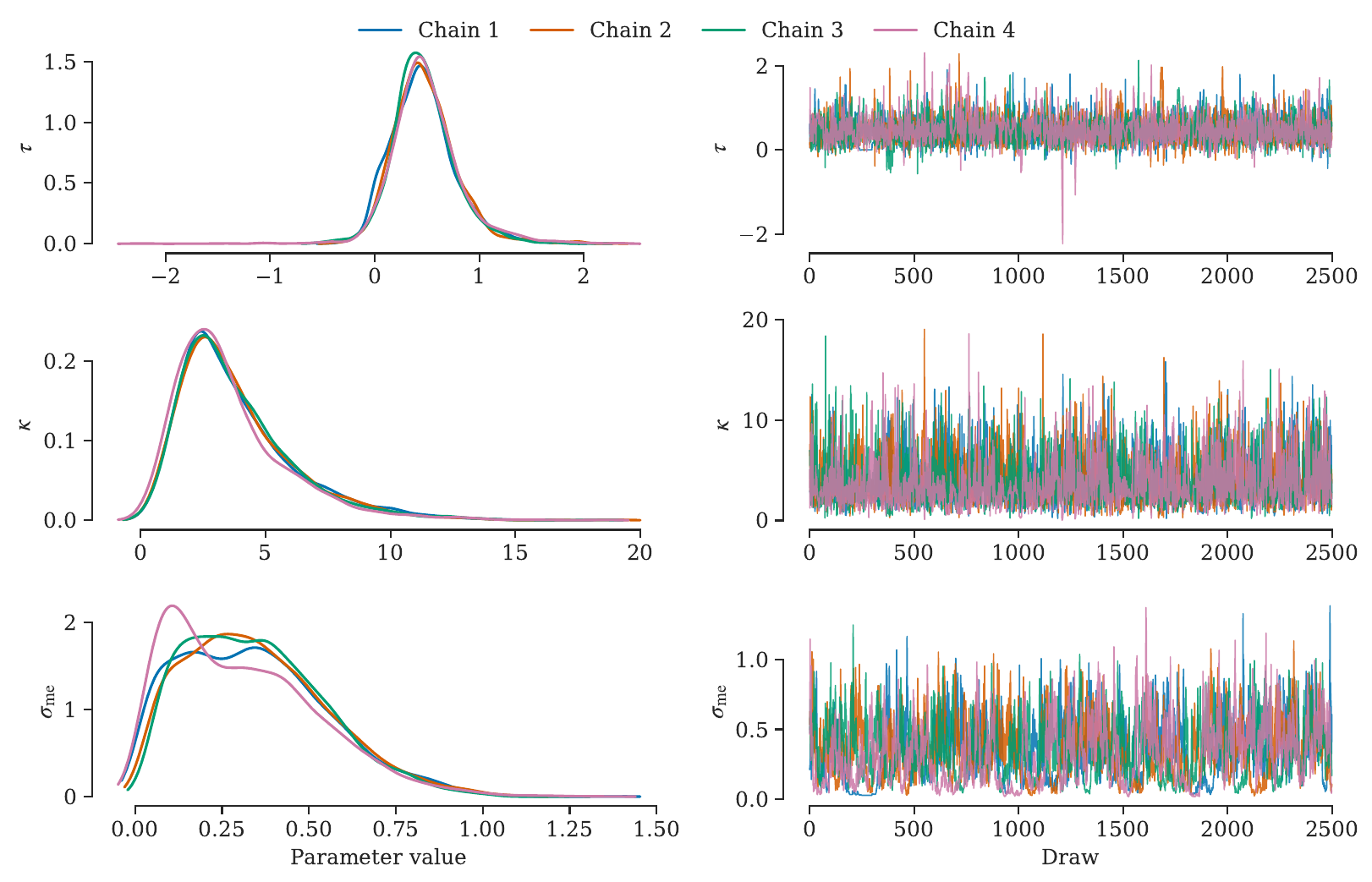}
\caption{MCMC diagnostic plots for the Bayesian soft-threshold model fitted to the MNHHS cohort, with parameters \((\tau, s, \sigma_{\mathrm{me}})\). For each parameter, the figure shows the chain-specific marginal posterior density together with the corresponding trace across sampling iterations.}
\label{fig:trace}
\end{figure}

\begin{figure}[H]
\centering
\includegraphics[width=0.92\linewidth]{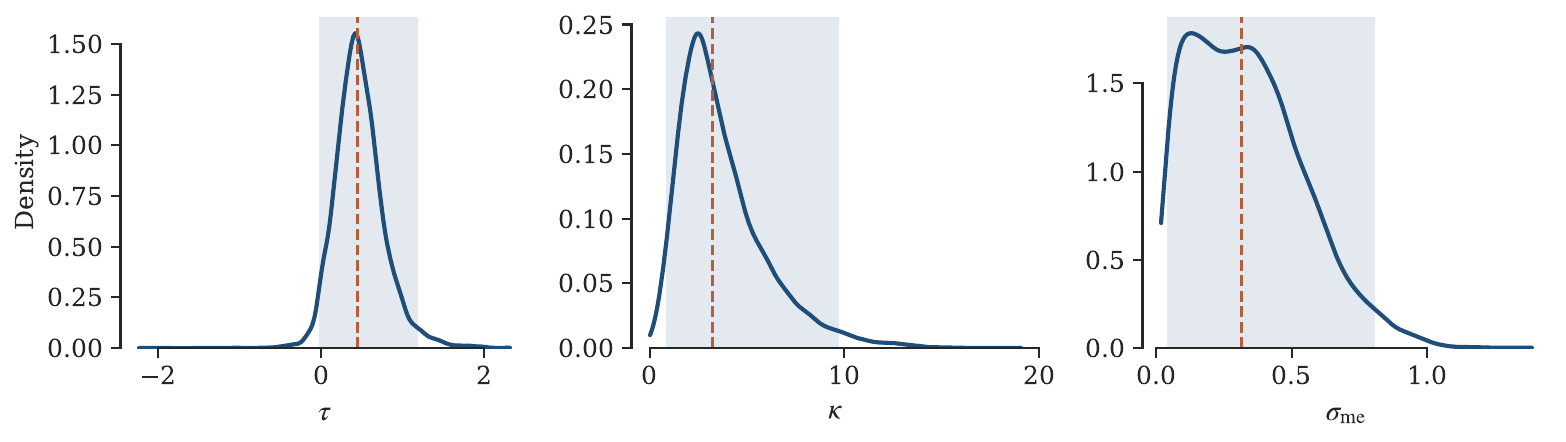}
\caption{Marginal posterior densities for the Bayesian soft-threshold model parameters \((\tau, s, \sigma_{\mathrm{me}})\) in the MNHHS cohort analysis. Shaded regions indicate the reported 95\% highest-density intervals, and dashed vertical lines indicate posterior medians.}
\label{fig:marginals}
\end{figure}

Supplementary Figures~\ref{fig:trace} and \ref{fig:marginals} reproduce the corresponding trace plots and marginal posterior densities.

\subsection{Posterior predictive summaries and performance metrics}
We report two complementary predictive summaries.

\paragraph{Full-fit analysis}
We fit the classifier once using all cases in the external cohort. For each case, we summarize the fitted growth probability by its posterior mean and use these probabilities to evaluate in-sample discrimination and calibration.

\paragraph{Pooled leave-one-out analysis}
For the MNHHS cohort, \(N=19\) aneurysms were available. We therefore carried out leave-one-out cross-validation over 19 folds, with each fold holding out one aneurysm for testing and fitting the model on the remaining \(N-1=18\) aneurysms. Within each fold, we recomputed the standardization parameters using only the corresponding training set so that the held-out aneurysm was always evaluated strictly out of sample. After completing all 19 folds, each aneurysm had one held-out posterior probability. We summarized the pooled leave-one-out analysis using the posterior \emph{median} probability for each held-out case.

We quantified discrimination with the area under the receiver operating characteristic curve (AUC). We assessed calibration with reliability diagrams and the Brier score. To measure agreement with the binary reference labels, we thresholded posterior growth probabilities at 0.5 and then computed percentage agreement and Cohen's \(\kappa\).

\section{Supplementary results}

\subsection{Reference-label prevalence}
Supplementary Table~\ref{tab:prevalence} reports the class prevalences used throughout the study. The institutional UCSF cohort shows lower apparent growth prevalence than the external MNHHS cohort, and the prevalence within UCSF depends on which rater serves as the reference. These differences are relevant because they influence both the operating difficulty of the classification task and the expected variability of calibration curves in modest-sized samples.

\begin{table}[H]
\centering
\caption{Class prevalence by cohort and reference label set.}
\label{tab:prevalence}
\begin{tabular}{@{}llcccc@{}}
\toprule
\textbf{Cohort} & \textbf{Reference label} & \textbf{\(n\)} & \textbf{Growing} & \textbf{Stable} & \textbf{Prevalence} \\
\midrule
UCSF  & Junior rater   & 42 & 11 & 31 & 26.2\% \\
UCSF  & Senior rater   & 42 &  8 & 34 & 19.0\% \\
MNHHS & External rater & 19 &  6 & 13 & 31.6\% \\
\bottomrule
\end{tabular}
\end{table}

\subsection{Posterior diagnostics for the external model}
The MCMC diagnostics for the MNHHS-fitted model indicate satisfactory posterior sampling behavior. Supplementary Figure~\ref{fig:trace} shows that the four chains mix without persistent drift or systematic chain separation. Supplementary Figure~\ref{fig:marginals} shows unimodal posterior marginals for the threshold, slope, and measurement-error parameters. No divergent transitions were reported in this run.

Supplementary Table~\ref{tab:posterior} summarizes the posterior distributions and associated MCMC diagnostics for the monitored parameters. All split-\(\hat R\) values were close to 1, with the largest observed for the measurement-error scale \(\sigma_{\mathrm{me}}\) (\(1.022\)), indicating satisfactory overall mixing. The bulk effective sample size was largest for the threshold parameter and smallest for \(\sigma_{\mathrm{me}}\), but remained sufficient to support stable posterior estimation. When mapped back to physical units, the threshold corresponds to a posterior median cut-off of \(0.133\) mm, with a 95\% equal-tail interval from \(0.076\) to \(0.221\) mm. Thus, the inferred 50\% transition point lies below the smallest voxel dimension (\(0.23\) mm), indicating that the model learns a decision threshold at a sub-voxel displacement scale.

\begin{table}[H]
\centering
\caption{Posterior summaries for the model fitted to the MNHHS cohort. Equal-tail 95\% posterior intervals are reported.}
\label{tab:posterior}
\small
\begin{tabular}{@{}lcccccc@{}}
\toprule
\textbf{Parameter} & \textbf{Mean} & \textbf{SD} & \textbf{Median} & \textbf{95\% interval} & \textbf{Split-\(\hat R\)} & \textbf{Bulk ESS} \\
\midrule
\(\tau\) (standardized)  & 0.476 & 0.308 & 0.449 & [-0.026, 1.187] & 1.002 & 2295 \\
\(\tau\) (mm)            & 0.136 & 0.037 & 0.133 & [0.076, 0.221]  & 1.002 & 2295 \\
\(s\)                    & 3.764 & 2.280 & 3.214 & [0.838, 9.734]  & 1.007 & 1237 \\
\(\sigma_{\mathrm{me}}\) & 0.337 & 0.208 & 0.316 & [0.039, 0.808]  & 1.022 & 248  \\
\bottomrule
\end{tabular}
\end{table}

\subsection{Posterior predictive validation}
Supplementary Table~\ref{tab:performance} summarizes predictive performance for the Bayesian soft-threshold growth model under both full-fit and leave-one-out evaluation on the MNHHS cohort. In the full-fit analysis, the model achieved an AUC of 0.872 and a Brier score of 0.119. The AUC indicates good discrimination between growth-positive and growth-negative aneurysms. The Brier score ranges from 0 to 1, with lower values indicating better probabilistic accuracy; here, the value of 0.119 indicates that the predicted probabilities remained reasonably close to the observed outcomes. Under pooled leave-one-out evaluation, the AUC decreased to 0.821 and the Brier score increased to 0.146. This modest deterioration is expected under strictly out-of-sample assessment, particularly given the small cohort size, and indicates that the model largely preserves both discrimination and probability accuracy.

\begin{table}[H]
\centering
\caption{Predictive performance of the Bayesian soft-threshold growth model on the MNHHS cohort under full-fit and leave-one-out evaluation. Full-fit metrics were computed from posterior mean growth probabilities, whereas pooled leave-one-out metrics were computed from held-out posterior median probabilities. AUC quantifies discrimination, and the Brier score quantifies probabilistic accuracy on a 0--1 scale, with lower values indicating better performance.}
\label{tab:performance}
\begin{tabular}{@{}lccccc@{}}
\toprule
\textbf{Analysis} & \textbf{\(n\)} & \textbf{AUC} & \textbf{Brier} & \textbf{Agreement} & \textbf{Cohen's \(\kappa\)} \\
\midrule
Full fit     & 19 & 0.872 & 0.119 & 78.9\% & 0.513 \\
Pooled LOOCV & 19 & 0.821 & 0.146 & 78.9\% & 0.513 \\
\bottomrule
\end{tabular}
\end{table}

Agreement with the binary reference labels was 78.9\% for both the full-fit and pooled leave-one-out analyses, with Cohen's \(\kappa=0.513\), indicating moderate agreement beyond chance.

\begin{figure}[H]
\centering
\begin{subfigure}[t]{0.47\linewidth}
    \centering
    \includegraphics[width=\linewidth]{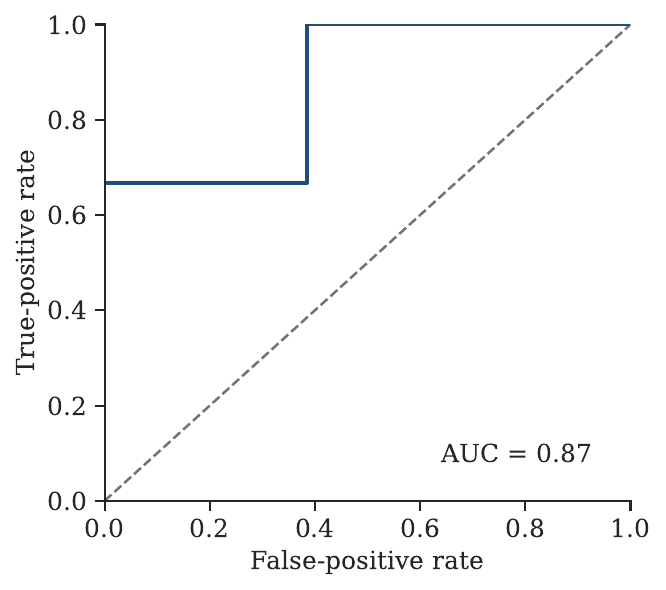}
    \caption{}
\end{subfigure}
\hfill
\begin{subfigure}[t]{0.47\linewidth}
    \centering
    \includegraphics[width=\linewidth]{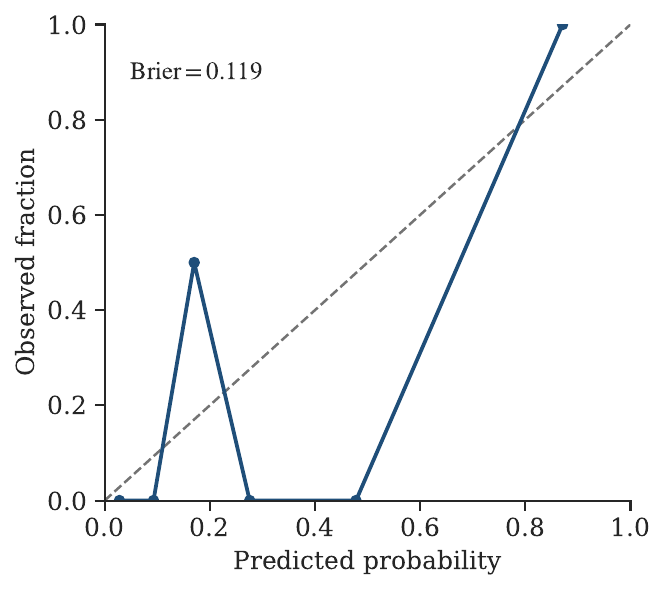}
    \caption{}
\end{subfigure}
\caption{Full-fit predictive evaluation of the Bayesian soft-threshold growth model on the MNHHS cohort. (a) Receiver operating characteristic curve based on posterior mean growth probabilities. (b) Reliability diagram for the same probabilities, comparing predicted growth risk with observed event frequency. The diagonal dashed line denotes ideal calibration.}
\label{fig:fullfit}
\end{figure}

\begin{figure}[H]
\centering
\begin{subfigure}[t]{0.47\linewidth}
    \centering
    \includegraphics[width=\linewidth]{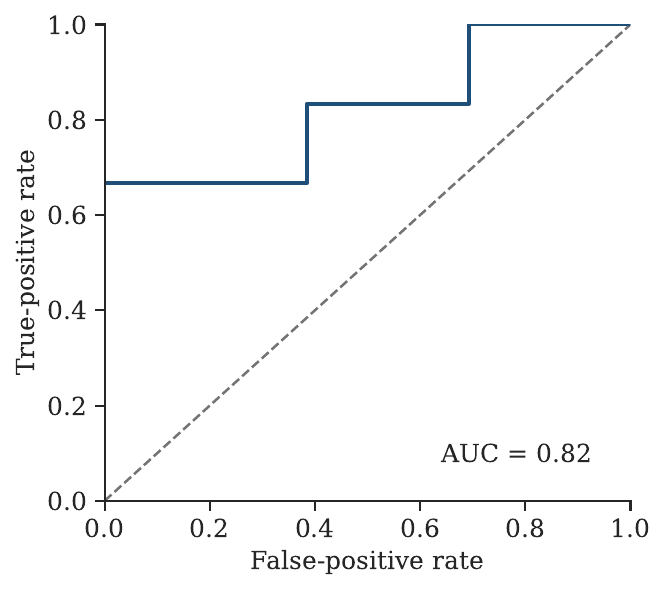}
    \caption{}
\end{subfigure}
\hfill
\begin{subfigure}[t]{0.47\linewidth}
    \centering
    \includegraphics[width=\linewidth]{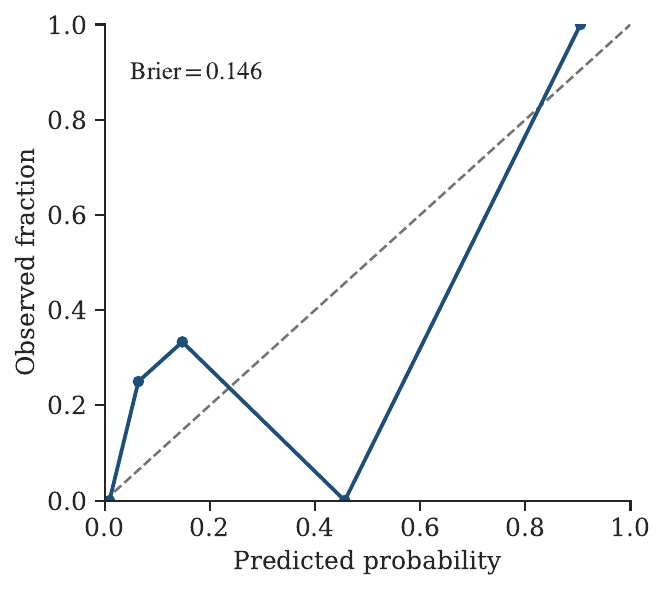}
    \caption{}
\end{subfigure}
\caption{Leave-one-out predictive evaluation of the Bayesian soft-threshold growth model on the MNHHS cohort. (a) Receiver operating characteristic curve based on pooled held-out posterior median probabilities. (b) Reliability diagram for the corresponding out-of-sample probabilities. The diagonal dashed line denotes ideal calibration.}
\label{fig:loo}
\end{figure}

Supplementary Figures~\ref{fig:fullfit} and \ref{fig:loo} provide ROC- and calibration-based summaries for the full-fit and leave-one-out analyses, respectively. We interpret the calibration plots qualitatively rather than as precise estimates because calibration assessment is unstable in a cohort of this size and becomes especially sensitive in the leave-one-out setting, where only six aneurysms were labeled as growth-positive. Even with this limitation, the plots remain consistent with the main finding that the model separates lower-probability from higher-probability cases in a clinically meaningful manner.

